\documentclass[final,3p,times,twocolumn,sort&compress]{elsarticle}

\usepackage{epsfig}
\usepackage{amssymb}
\usepackage{amsthm}
\usepackage{amsmath}
\usepackage{color}
\usepackage{graphicx}
\usepackage[table]{xcolor}
\usepackage{lscape}
\usepackage{caption}
\usepackage{longtable}
\usepackage{soul}

\journal{Journal of Magnetism and Magnetic Materials}

\begin{document}
\setstcolor{red}
\begin{frontmatter}



\title{Magnetization processes and existence of reentrant phase transitions in coupled spin-electron model on doubly decorated planar lattices}


\author{Hana \v Cen\v carikov\'a\corref{cor1}\fnref{label1}}
\cortext[cor1]{Corresponding author:}
\ead{hcencar@saske.sk}
\author{Jozef Stre\v{c}ka\fnref{label2}}
\author{Andrej Gendiar\fnref{label3}}

\address[label1]{Institute of Experimental Physics, Slovak Academy of Sciences, Watsonova 47, 040 01 Ko\v {s}ice, Slovakia}
\address[label2]{Department of Theoretical Physics and Astrophysics, Faculty of Science, P.~J. \v{S}af\' arik University, Park Angelinum 9, 040 01 Ko\v{s}ice, Slovakia}
\address[label3]{Institute of Physics, Slovak Academy of Sciences, D\'{u}bravsk\'{a} cesta 9, SK-845 11, Bratislava, Slovakia}
\begin{abstract}
An alternative model for a description of magnetization processes in coupled 2D spin-electron systems has been introduced and rigorously examined using the generalized decoration-iteration transformation and the corner transfer matrix renormalization group method. The model consists of localized Ising spins placed on nodal lattice sites and mobile electrons delocalized over the pairs of decorating sites. It takes into account a hopping term for mobile electrons, the Ising coupling between mobile electrons and localized spins as well as the Zeeman term acting on both types of particles. The ground-state and finite-temperature phase diagrams were established and comprehensively analyzed. It was found that the ground-state phase diagrams are very rich depending on the electron hopping and applied magnetic field. The diversity of magnetization curves can be related to intermediate magnetization plateaus, which may be continuously tuned through the density of mobile electrons. In addition, the existence of several types of reentrant phase transitions driven either by temperature or magnetic field was proven. 
\end{abstract}


\end{frontmatter}


\section{Introduction}
\label{s1}
Correlated spin-electron systems belong to the intensively studied materials in the condensed matter physics due to the variety of unconventional structural, electronic, magnetic, and transport properties~\cite{Kanamori,Takada,Honecker,Kikuchi,Kamihara1,Kamihara2,Li,Koster} implying wide application potential. An exhaustive understanding of their origins thus opens a new way in technological applications, where suitable and unconventional properties could be used simultaneously.

To achieve this goal, various types of models~\cite{Imada,Dagotto2,Schwieger,Cenci3} in combination with less or more sophisticated methods~\cite{Filippetti,Schollwock,ALPS,Haerter,Cenci4} have been used. In spite of enormous effort, some of physical phenomena still lack full understanding since they have not been reliably explained so far. In the present paper we will consider two-dimensional (2D) coupled spin-electron systems consisting of localized Ising spins and mobile electrons using a relatively simple analytical method based on the Fisher mapping idea~\cite{Fisher}. In this concept an arbitrary statistical-mechanical system, which merely interacts either with two or three outer Ising spins, may be replaced with effective interactions between the outer Ising spins through the generalized decoration-iteration or star-triangle mapping transformations~\cite{Fisher,Syozi,Rojas}. This procedure was successfully applied to simulate magnetic properties of various two-component spin-electron systems in one~\cite{Pereira1,Pereira2,Cisarova,Cisarova2,Cisarova3,Galisova3} or two dimensions~\cite{Galisova,Galisova2,Doria,Cenci,Cenci2} with a good qualitative coincidence of magnetic behavior in real materials. For example, a 2D coupled spin-electron model could be used as a simplified theoretical model of selected rare-earth compounds, manganites or intermetallics with a quasi-2D character in which the existence of metamagnetic or reentrant transitions was observed~\cite{Venturini,Kolmakova,Ghimire,Mihalik}. These unconventional magnetic phenomena put aforementioned two-component spin-electron systems into the center of research interest, because one may easily control their magnetic states and thus manage various processes driven through a relatively simple change of external parameters, as for instance, temperature and/or magnetic field.

In the present work we concentrate our attention to a coupled spin-electron model on the square lattice, which contains the localized Ising spins situated at the nodal lattice sites while mobile electrons are delocalized over the pairs of decorating sites placed at each of its bonds. Our previous study of the identical model with the absence of external magnetic field~\cite{Cenci,Cenci2} has shown that this relatively simple model is able to describe a rich spectrum of unconventional physical properties, like the existence of various magnetic phases or presence of interesting reentrant phase transition in both the ferromagnetic and the antiferomagnetic limit. The rich spectrum of the zero-field model has motivated us to inspect the behavior of the same model under the influence of external magnetic field with the main goal to examine the existence of magnetization plateaus accompanied by the presence of metamagnetic transitions. Besides, we also concentrate on the existence of reentrant phase transitions in order to observe influence of external magnetic field and temperature on phase-transition stability. 

The organization of the paper is as follows. In Section~\ref{s2} we will introduce the investigated model and the method used for its solution. The ground-state and finite-temperature phase diagrams will be discussed in detail in Section~\ref{results} along with the thermal dependencies of magnetization. In this section, we focus on possible emergence of the reentrant phase transitions as well as the intermediate magnetization plateaus in magnetization curves. Finally, the most significant results will be summarized together with future outlooks in the Section~\ref{Conclusions}. 

\section{Model and Method}
\label{s2}

\begin{figure}[h!]
\begin{center}
\includegraphics[scale=0.9]{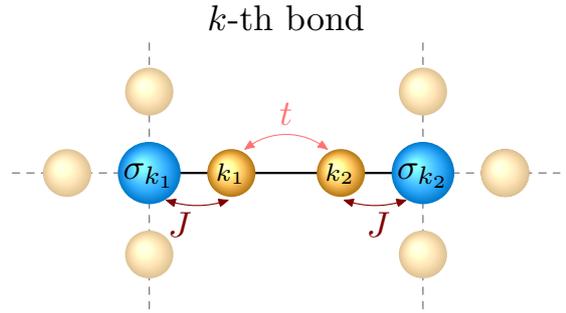}
\caption{\small A schematic representation of the $k$-th bond in the studied spin-electron model (\ref{eq1}) on the doubly decorated square lattice. Bigger balls correspond to the nodal lattice sites occupied by the localized Ising spins, while smaller balls to the decorating sites occupied by at most four mobile electrons per dimer. The interactions assumed within the $k$-th bond are visualized.}
\label{fig1}
\end{center}
\end{figure}

Let us define an interacting spin-electron system on doubly decorated 2D lattices. The investigated model contains one localized Ising spin at each nodal lattice site and a set of mobile electrons delocalized over the pairs of decorating sites (dimers) placed at each bond. The total Hamiltonian of investigated spin-electron model  can be written as a sum over all  bond Hamiltonians $\hat{\cal H}=\sum_{k=1}^{Nq/2} \hat{\cal H}_k$, where the symbol $q$ denotes the coordination number of the underlying 2D lattice and $N$ denotes the total number of its nodal lattice sites. Each bond Hamiltonian $\hat{\cal H}_k$ contains the kinetic energy of mobile electrons on the $k$-th bond, the exchange interaction between the mobile electrons and their nearest-neighbor Ising spins, and the Zeeman terms describing the influence of external magnetic field on the magnetic moment of mobile electrons and localized spins (see Fig. \ref{fig1})
\begin{align}\hspace*{-0.4cm}
\hat{\cal H}_k=&-t(\hat{c}^\dagger_{k_1,\uparrow}\hat{c}_{k_2,\uparrow}^{~}+\hat{c}^\dagger_{k_1,\downarrow}\hat{c}_{k_2,\downarrow}^{~}+
\hat{c}^\dagger_{k_2,\uparrow}\hat{c}_{k_1,\uparrow}^{~}+\hat{c}^\dagger_{k_2,\downarrow}\hat{c}_{k_1,\downarrow}^{~})
\nonumber
\\
&-J\hat\sigma^z_{k_1}(\hat{n}_{k_1,\uparrow}-\hat{n}_{k_1,\downarrow})-
J\hat\sigma^z_{k_2}(\hat{n}_{k_2,\uparrow}-\hat{n}_{k_2,\downarrow})
\nonumber\\
&-h_e(\hat{n}_{k_1,\uparrow}-\hat{n}_{k_1,\downarrow})-
h_e(\hat{n}_{k_2,\uparrow}-\hat{n}_{k_2,\downarrow})
\nonumber\\
&-\frac{h_i}{q}(\hat\sigma^z_{k_1}+\hat\sigma^z_{k_2})\;.
\label{eq1}
\end{align}
The symbols $\hat{c}^\dagger_{k_{\alpha},\gamma}$ and $\hat{c}_{k_{\alpha},\gamma}^{~}$ ($\alpha=1,2$; $\gamma=\uparrow,\downarrow$) in Eq.~(\ref{eq1}) represent the creation and annihilation fermionic operators for the mobile electron. The respective number operators are denoted by $\hat{n}_{k_{\alpha},\gamma}=\hat{c}^\dagger_{k_{\alpha},\gamma}\hat{c}_{k_{\alpha},\gamma}^{~}$ and $\hat{n}_{k_{\alpha}}=\hat{n}_{k_{\alpha},\uparrow}+\hat{n}_{k_{\alpha},\downarrow}$. $\hat\sigma^z_{k_{\alpha}}$ denotes the $z$-component of the Pauli operator with the eigenvalues $\sigma=\pm1$. The first term in Eq.~(\ref{eq1}) corresponds to the kinetic energy of mobile electrons delocalized over a couple of decorating  sites $k_1$ and $k_2$ from the $k$-th dimer modulated by the hopping amplitude $t$. The second and the third terms represent the Ising interaction between the mobile electrons and their nearest-neighbor Ising spins described by the parameter $J$. Finally, the last three terms in Eq.~(\ref{eq1}) correspond to the Zeeman energy of the  magnetic moments relevant to  the localized spins ($h_i$) and the delocalized electrons ($h_e$). The dissimilarity between the magnetic fields is implemented for the purpose of analytic calculations only, however, both the magnetic fields are considered equal to one another, $h_i=h_e=h$ at the final stage of our analysis.  For illustration, a schematic representation of the $k$-th bond of the aforementioned model is displayed in Fig.~\ref{fig1} for a special case of the doubly decorated square lattice. It should be noted, however, that all derivations presented in Section~\ref{s2} are general and hold for an arbitrary 2D lattice.

To study the ground-state as well as thermodynamic properties of the coupled spin-electron system defined by the Hamiltonian (\ref{eq1}), it is necessary to consider the grand-canonical partition function $\Xi$
\begin{align}
\Xi=\sum_{\{\sigma\}}\mbox{Tr}\exp\left[-\beta \sum_{k=1}^{Nq/2} \left(\hat{\cal H}_k
   - \mu\hat{n}_k\right) \right] \,,
\label{eq2}
\end{align}
where $\beta=1/(k_BT)$, $k_B$ is the Boltzmann's constant, $T$ is the absolute temperature, $\hat{n}_k=\hat{n}_{k_1}+\hat{n}_{k_2}$ is the number operator of mobile electrons delocalized over the $k$-th decorating dimer and $\mu$ is the chemical potential. The summation in Eq.~(\ref{eq2}) runs over all possible spin configurations $\{\sigma\}$ of the nodal Ising spins and the symbol Tr stands for the trace over the degrees of freedom of the mobile electrons only. Assuming the mutual commutativity of two bond Hamiltonians, i.e. $[\hat{\cal H}_i,\hat{\cal H}_j]=0$, one can partially factorize the grand-canonical partition function into the product of bond partition functions $\Xi_k$ 
\begin{align}
\Xi&=\sum_{\{\sigma\}}\prod_{k=1}^{Nq/2}\mbox{Tr}_k\exp(-\beta\hat{\cal H}_k)
\exp(\beta\mu\hat{n}_k)\nonumber\\
&=\sum_{\{\sigma\}}\prod_{k=1}^{Nq/2}\Xi_k\;.
\label{eq3}
\end{align}
Here, the symbol Tr$_k$ stands for the trace over the degrees of freedom of the mobile electrons from the $k$-th decorating dimer. This simplification allows us to calculate the partition function $\Xi$ exactly out of the eigenvalues of the bond Hamiltonian (\ref{eq1}). Validity of the commutative relation between the bond Hamiltonians $\hat{\cal H}_k$ and the number operator of mobile electrons per bond ($\hat{n}_k$) implies that the matrix form of the bond Hamiltonian $\hat{\cal H}_k$ can be divided into several disjoint blocks ${\cal H}_k(n_k)$ corresponding to the respective orthogonal Hilbert subspaces, which are characterized by different number of the mobile electrons ($n_k$) per bond. Thus, the eigenvalues of the bond Hamiltonians $\hat{\cal H}_k$ can be calculated straightforwardly
\begin{eqnarray}\hspace*{-0.9cm}
\begin{array}{ll}
n_k=0:& E_{k_1}=-h_iL/q\,,\\
n_k=1:& E_{k_2}= -JL/2+\sqrt{J^2P^2+4t^2}/2-h_iL/q-h_e\,,\\
& E_{k_3}=-JL/2-\sqrt{J^2P^2+4t^2}/2-h_iL/q-h_e\,,\\
& E_{k_4}=+JL/2+\sqrt{J^2P^2+4t^2}/2-h_iL/q+h_e\,,\\
& E_{k_5}=+JL/2-\sqrt{J^2P^2+4t^2}/2-h_iL/q+h_e\,,\\
n_k=2: & E_{k_6}=-JL-h_iL/q-2h_e\,,\\
& E_{k_7}=+JL-h_iL/q+2h_e\,,\\
 & E_{k_8}=E_{k_9}=-h_iL/q\,,\\
 & E_{k_{10}}=+\sqrt{J^2P^2+4t^2}-h_iL/q\,,\\
  & E_{k_{11}}=-\sqrt{J^2P^2+4t^2}-h_iL/q\,,\\
n_k=3: & E_{k_{12}}= -JL/2+\sqrt{J^2P^2+4t^2}/2-h_iL/q-h_e\,,\\
& E_{k_{13}}=-JL/2-\sqrt{J^2P^2+4t^2}/2-h_iL/q-h_e\,,\\
& E_{k_{14}}=+JL/2+\sqrt{J^2P^2+4t^2}/2-h_iL/q+h_e\,,\\
& E_{k_{15}}=+JL/2-\sqrt{J^2P^2+4t^2}/2-h_iL/q+h_e\,,\\
n_k=4: &E_{k_{16}}=-h_iL/q\,.
\end{array}
\hspace{-0.9cm}\label{eq4}
\end{eqnarray}
For simplification, we have defined here two parameters $L=\sigma_{k_1}+\sigma_{k_2}$ and $P=\sigma_{k_1}-\sigma_{k_2}$.

After tracing out the degrees of freedom of mobile electrons, the bond grand-canonical partition function $\Xi_k$ depends only on the spin states of two localized Ising spins, whereas its explicit form can be replaced with the generalized decoration-iteration transformation~\cite{Fisher,Syozi,Rojas} 
\begin{subequations}
\begin{align}
\Xi_k&=\sum_{i=1}^{16}\exp(-\beta E_{k_i})\exp\left[\beta\mu n_k(E_{k_i})\right]
\nonumber\\
&=\exp\left(\frac{\beta h_iL}{q}\right)\bigg\{1+4(z+z^3)\cosh\left[\beta\left(\frac{JL}{2}+h_e\right)\right]\times
\nonumber\\
&\cosh\left[\frac{\beta}{2}\sqrt{J^2P^2+4t^2}\right]
+ 2z^2\left\{ 1+\cosh\left[\beta (JL+h_e)\right]\right.
\nonumber\\
&\left.+\cosh\left[\beta\sqrt{J^2P^2+4t^2}\right]\right\} + z^4  \bigg\}
\label{eq16a} \\
&= A\exp(\beta R\sigma_{k_1}\sigma_{k_2})\exp\left(\frac{\beta h_{ef}L}{q}\right)\;.
\label{eq16b}
\end{align}
\label{eq16}
\end{subequations}
Here, $z=\exp(\beta\mu)$ is used to denote the fugacity of the mobile electrons. The physical meaning of this decoration-iteration transformation (\ref{eq16}) lies in replacement of a more complicated system (\ref{eq16a}) by its simpler counterpart (\ref{eq16b}) with new effective interactions. The evaluation of the mapping parameters $A$, $R$, and $h_{ef}$ are given by "self-consistent" condition of the decoration-iteration transformation (\ref{eq16}), which must hold for all four combinations of the two Ising spins $\sigma_{k_1}$ and $\sigma_{k_2}$ requiring
\begin{align}
&A=(V_1V_2V_3^2)^{1/4},\hspace*{1cm} \beta R=\frac{1}{4}\ln\left(\frac{V_1V_2}{V_3^2}\right),
\nonumber\\
&\beta h_{ef}=\frac{q}{4}\ln\left(\frac{V_1}{V_2}\right), 
\label{eq17}
\end{align}
where 
\begin{align}
V_1&=\exp\left(\frac{2\beta h_i}{q}\right)\left\{1+z^4+4(z+z^3)\cosh\left[\beta(J+h_e)\right]\times\right.
\nonumber\\
&\left.\cosh(\beta t)+ 2z^2\left[ 1+\cosh\left[2\beta(J+h_e)\right]
+\cosh(2\beta t)\right]\right\}\; ,
\nonumber\\
V_2&=\exp\left(\frac{-2\beta h_i}
{q}\right)\left\{1+z^4+4(z+z^3)\cosh\left[\beta(J-h_e)\right]\times\right.
\nonumber\\
&\left.\cosh(\beta t)+2z^2\left[ 1+\cosh\left[2\beta(J-h_e)\right]
+\cosh(2\beta t)\right]\right\}\; ,
\nonumber\\
V_3&=1+z^4+4(z+z^3)\cosh\left(\beta\sqrt{J^2+t^2}\right)\cosh(\beta h_e)
\nonumber\\
&+2z^2\left[ 1+\cosh\left(2\beta\sqrt{J^2+t^2}\right)+\cosh(2\beta h_e)\right]\;.
\label{eq18}
\end{align}
Substituting  the transformation (\ref{eq16a}) - (\ref{eq16b}) into the expression (\ref{eq3}), one obtains a simple mapping relation between the grand-canonical partition function $\Xi$ of the interacting spin-electron system on the doubly decorated 2D lattices and, respectively, the canonical partition function $Z_{IM}$ of a simple Ising model on the corresponding undecorated lattice with an effective nearest-neighbor interaction $R$ and effective field $h_{ef}$
\begin{align}
\Xi(\beta,J,t,h)=A^{Nq/2}Z_{IM}(\beta,R,h_{ef})\;.
\label{eq19}
\end{align}
Obviously, the mapping parameter $A$ cannot cause non-analytic behavior of the grand-canonical partition function $\Xi$. Hence, the investigated spin-electron system becomes critical if and only if the corresponding Ising model becomes critical as well.

To study the model behavior in context of various electron concentrations, it is necessary to determine the equation of state relating its mean value $\langle n_k\rangle$ with respect to the model parameters. The mean electron concentration $\langle n_k\rangle$ can be straightforwardly derived from the grand potential $\Omega=-k_BT\ln \Xi$
\begin{align}
\rho&\equiv\langle n_k\rangle=-\left(\frac{\partial \Omega}{\partial \mu}\right)_T=\frac{z}{Nq/2}\frac{\partial}{\partial z}\ln\Xi\nonumber\\
&=z\frac{\partial}{\partial z}\ln A+z\varepsilon\frac{\partial}{\partial z}\beta R
+\frac{z}{q}\langle\sigma_{k_1}+\sigma_{k_2}\rangle\frac{\partial}{\partial z}\beta h_{ef}
\nonumber\\
&=\frac{z}{4}\left(\frac{V'_1}{V_1}+\frac{V'_2}{V_2}+2\frac{V'_3}{V_3}\right)+\frac{z}{4}\varepsilon\left(\frac{V'_1}{V_1}+\frac{V'_2}{V_2}-2\frac{V'_3}{V_3}\right)
\nonumber\\
&+\frac{z}{2}\langle \sigma_{k_1}\rangle\left(\frac{V'_1}{V_1}-\frac{V'_2}{V_2}\right)
\;,
\label{eq20}
\end{align}
where $\varepsilon=\langle\sigma_{k_1}\sigma_{k_2}\rangle$ denotes the nearest-neighbor pair correlation function and
\begin{align}
V'_1&=\frac{\partial V_1}{\partial z}=4\exp(2\beta h_i/q)\left\{(1+3z^2)\cosh\left(\beta (J+h_e)\right)\times\right.
\nonumber\\
&\left.\cosh(\beta t)+z\left[1+\cosh\left(2\beta (J+h_e)\right)+\cosh\left(2\beta t\right)\right]\right.
\nonumber\\
&\left.+z^3\right\}\;,
\nonumber\\
V'_2&=\frac{\partial V_2}{\partial z}=4\exp(-2\beta h_i/q)\left\{(1+3z^2)\cosh\left(\beta (J-h_e)\right)\times\right.
\nonumber\\
&\left.\cosh(\beta t)+z\left[1+\cosh\left(2\beta (J-h_e)\right)+\cosh\left(2\beta t\right)\right]\right.
\nonumber\\
&\left.+z^3\right\}\;,
\nonumber\\
V'_3&=\frac{\partial V_3}{\partial z}=4(1+3z^2)\cosh\left(\beta\sqrt{J^2+t^2}\right)\cosh\left(\beta h_e\right)
\nonumber\\
&+4z\left[1+\cosh\left(2\beta\sqrt{J^2+t^2}\right)+\cosh\left(2\beta h_e\right)\right]+4z^3\;.
\label{eq21}
\end{align}
As mentioned above, the main goal of this paper is to analyze the magnetization processes of the coupled 2D spin-electron model, and for this purpose, we separately derive expressions for the uniform sublattice magnetizations of localized spins $m_i$ and the mobile electrons $m_e$ per elementary unit cell
\begin{align}
m_i=-\left(\frac{\partial \Omega}{\partial h_i}\right)_{z},\hspace{1.cm}m_e=-\left(\frac{\partial \Omega}{\partial h_e}\right)_{z}.
\label{eq21a}
\end{align}
The final formulas for the uniform sublattice magnetizations relate with the partial derivatives of all mapping parameters with respect to the relevant local fields $h_i$ and $h_e$
\begin{align}
m_i&=\frac{q}{2}\left[\frac{\partial \ln A}{\partial \beta h_i}+\varepsilon\frac{\partial \beta R}{\partial \beta h_i} +\frac{2m_{IM}}{q}\frac{\partial \beta h_{ef}}{\partial \beta h_i}\right]=m_{IM}\\
m_e&=\frac{q}{2}\left[\frac{\partial \ln A}{\partial \beta h_e}+\varepsilon\frac{\partial \beta R}{\partial \beta h_e} +\frac{2m_{IM}}{q}\frac{\partial \beta h_{ef}}{\partial \beta h_e}\right]\nonumber\\
&=\frac{q}{8}\left[(1+\varepsilon) \left(\frac{W_1}{V_1}+\frac{W_2}{V_2}\right)+2(1-\varepsilon)\frac{W_3}{V_3}\right.
\nonumber\\
&+\left.2m_{IM}\left(\frac{W_1}{V_1}-\frac{W_2}{V_2}\right)\right],
\label{eq21b}
\end{align}
where the coefficients $W_1$, $W_2$ and $W_3$ are defined as follows:
\begin{align}
W_1&=\frac{\partial V_1}{\partial \beta h_e}=4 \exp{(2\beta h_i/q)}\left[ (z+z^3)\cosh(\beta t)\times\right.
\nonumber\\
&\left.\sinh(\beta (J+h_e))+z^2\sinh(2\beta (J+h_e))\right]\;,\\
W_2&=\frac{\partial V_2}{\partial \beta h_e}=-4 \exp{(-2\beta h_i/q)}\left[ (z+z^3)\cosh(\beta t)\times\right.
\nonumber\\
&\left.\sinh(\beta (J-h_e))+z^2\sinh(2\beta (J-h_e))\right]\;,\\
W_3&=\frac{\partial V_3}{\partial \beta h_e}=4 \left[ (z+z^3)\cosh(\beta \sqrt{J^2+t^2})\sinh(\beta (h_e))\right.
\nonumber\\
&+\left.z^2\sinh(2\beta (h_e))\right].
\label{eq21c}
\end{align}
The total uniform magnetization of the coupled spin-electron model normalized with respect to its saturation value is then given by
\begin{align}
m_{tot}=\frac{m_i+m_e}{1+\rho}.
\label{eq21d}
\end{align}
It should be mentioned that the uniform magnetization is a convenient order parameter of the ferromagnetic ordering, however, it is inapplicable to the antiferromagnetic type of ordering. For this reason, we define new order parameters known as the staggered  sublattice magnetizations of localized spins $m^s_i$ and the mobile electrons $m^s_e$ 
\begin{align}
m^s_{i}&=\frac{1}{2}\langle \sigma_{k_1}-\sigma_{k_2}\rangle=m^s_{IM},
\label{eq21e}
\\
m^s_{e}&=\Bigg\langle \frac{1}{\Xi_k}\left[\frac{\partial \Xi_k}{\partial \beta J\sigma_{k_1}}-\frac{\partial \Xi_k}{\partial \beta J\sigma_{k_2}}\right]\Bigg\rangle
\nonumber\\
&=\frac{4Jm^s_i}{V_3\sqrt{J^2+t^2}}\left[(z+z^3)\sinh(\beta\sqrt{J^2+t^2})\cosh(\beta h_e)\right.
\nonumber\\
&+\left.z^2\sinh(2\beta\sqrt{J^2+t^2})\right],
\label{eq21f}
\end{align}
and, in analogy to the former case, the total staggered magnetization normalized to its saturation value is defined
\begin{align}
m^s_{tot}=\frac{m_i^s+m_e^s}{1+\rho}.
\label{eq21g}
\end{align}

All the derived analytical expressions depend on the uniform (\ref{eq21d}) and the staggered (\ref{eq21g}) magnetizations of the effective Ising model as well as on the nearest-neighbor correlation function $\varepsilon$, cf. Eqs. (\ref{eq20}) and (\ref{eq21b}). To evaluate those quantities accurately, we have adapted the Corner Transfer Matrix Renormalization Group (CTMRG) method~\cite{Nashino} to all the subsequent calculations. The CTMRG is a numerical algorithm applicable to 2D classical lattice spin models, which is build on ideas of the Density Matrix Renormalization Group method~\cite{White}. It enables to calculate all the thermodynamic functions efficiently and accurately. The main advantage of CTMRG lies in higher numerical accuracy of the thermodynamic functions (if compared with the Monte Carlo simulations~\cite{Binder}) when analyzing phase transitions and their critical behavior in various 2D spin systems.

\section{Results and discussion}
\label{results}
The following section introduces the most interesting results obtained from the study of the magnetization processes in the coupled spin-electron model on doubly decorated square lattice with the coordination number $q=4$  and the ferromagnetic nearest-neighbor interaction $J>0$ between the localized spins and mobile electrons. Without the loss of generality, the magnitude of this interaction and the Boltzmann constant are set to unity (i.e., $J=1$, $k_B=1$). The number of the free parameters can be further reduced by considering $h_i=h_e=h>0$. Finally, the electron density per decorating dimer may be restricted up to the half filling at most, i.e. $0\leq\rho\leq 2$ since the particle-hole symmetry applies for $\rho>2$.
\subsection{Ground state}
We start our discussion with the ground-state phase diagrams established in the $\mu$-$h$ plane for a wide range of the hopping parameter $t$. The detailed description of all possible phases forming the ground-state phase diagrams is listed in Tab.~\ref{tab1} along with the associated ground-state energies $E$. 
\begin{table*}[t!]
\begin{center}
\resizebox{1\textwidth}{!} {
\begin{tabular}{c||l|l}
Electron filling & Eigenvalue (${E})$& Eigenvector \\
\hline\hline
 $\begin{array}{c}
    \rho=0
   \end{array}$       
&
$\displaystyle
\begin{array}{l}
    {E}(\mbox{0})=-2h/q\\
 \end{array}   $
&
$\displaystyle
\begin{array}{l}
   |\mbox{0}\rangle=\prod_{k=1}^{Nq/2}|1\rangle_{\sigma_{k_1}}\otimes |0,0\rangle_{k}\otimes|1\rangle_{\sigma_{k_2}}\\
  \end{array}   $
\\
\hline
 $\begin{array}{c}
    \rho=1
   \end{array}$       
&
$\displaystyle
\begin{array}{l}
    {E}(\mbox{I})=-J-h-2h/q-t-\mu
   \end{array}   $
&
$\displaystyle
\begin{array}{l}
   |\mbox{I}\rangle=\prod_{k=1}^{Nq/2}|1\rangle_{\sigma_{k_1}}\otimes \frac{1}{\sqrt{2}}\left(
|\!\uparrow,0\rangle_k+|0,\uparrow\rangle_k\right)\otimes|1\rangle_{\sigma_{k_2}}
  \end{array}   $
 \\
\hline
 $\begin{array}{c}
    \rho=2
   \end{array}$       
&
$\displaystyle
\begin{array}{l}
    {E}(\mbox{II}_1)=-2J-2h-2h/q-2\mu\\
    {E}(\mbox{II}_2)=-2h/q-2t-2\mu\\
    {E}(\mbox{II}_3)=-2\sqrt{J^2+t^2}-2\mu\\
\end{array}   $
&
$\displaystyle
\begin{array}{l}
    |\mbox{II$_1$}\rangle=\prod_{k=1}^{Nq/2}|1\rangle_{\sigma_{k_1}}\otimes |\!\uparrow,\uparrow\rangle_k\otimes|1\rangle_{\sigma_{k_2}}\\
|\mbox{II}_2\rangle=\prod_{k=1}^{Nq/2}|1\rangle_{\sigma_{k_1}}\otimes\frac{1}{2}\left[|\!\uparrow,\downarrow\rangle_k-|\!\downarrow,\uparrow\rangle_k
+|\!\uparrow\downarrow,0\rangle_k+|0,\uparrow\downarrow\rangle_k\right]\otimes|1\rangle_{\sigma_{k_2}}    \\
 |\mbox{II$_3$}\rangle=\prod_{k=1}^{Nq/2}|1\rangle_{\sigma_{k_1}}\otimes \left[a|\!\uparrow,\downarrow\rangle_k+b|\!\downarrow,\uparrow\rangle_k
+c\left(|\!\uparrow\downarrow,0\rangle_k+|0,\uparrow\downarrow\rangle_k\right)\right]\otimes|-1\rangle_{\sigma_{k_2}}  
   \end{array}  $
\\
\hline
 $\begin{array}{c}
    \rho=3
   \end{array}$       
&
$\displaystyle
\begin{array}{l}
    {E}(\mbox{III})=-J-h-2h/q-t-3\mu
  \end{array}   $
&
$\displaystyle
\begin{array}{l}
   |\mbox{III}\rangle=\prod_{k=1}^{Nq/2}|1\rangle_{\sigma_{k_1}}\otimes \frac{1}{\sqrt{2}}\left(|\!\uparrow\downarrow,\uparrow\rangle_k-|\!\uparrow,\uparrow\downarrow\rangle_k\right)\otimes|1\rangle_{\sigma_{k_2}}
  \end{array}   $
 \\
\hline
 $\begin{array}{c}
    \rho=4
   \end{array}$       
&
$\displaystyle
\begin{array}{l}
    {E}(\mbox{IV})=-2h/q-4\mu\\
 \end{array}   $
&
$\displaystyle
\begin{array}{l}
   |\mbox{IV}\rangle=\prod_{k=1}^{Nq/2}|1\rangle_{\sigma_{k_1}}\otimes |\!\uparrow\downarrow,\uparrow\downarrow\rangle_k\otimes|1\rangle_{\sigma_{k_2}} \\
  \end{array}   $
 \\
\hline
\end{tabular}}
\end{center}
\captionof{table}{The list of eigenvalues and eigenvectors forming individual ground states. The probability amplitudes $a$, $b$, and $c$ used in the notation of the eigenvector  $|$II$_3\rangle$ have the following explicit form: $a=\frac{J+\sqrt{J^2+t^2}}{2\sqrt{J^2+t^2}}$, $b=\frac{-(\sqrt{J^2+t^2}-J)}{2\sqrt{J^2+t^2}}$, and $c=\frac{t}{2\sqrt{J^2+t^2}}$. }
\label{tab1}
\end{table*}   

\begin{figure}[!]
\begin{center}
\includegraphics[width=0.4\textwidth,trim=0 0 1.3cm 0.5cm, clip]{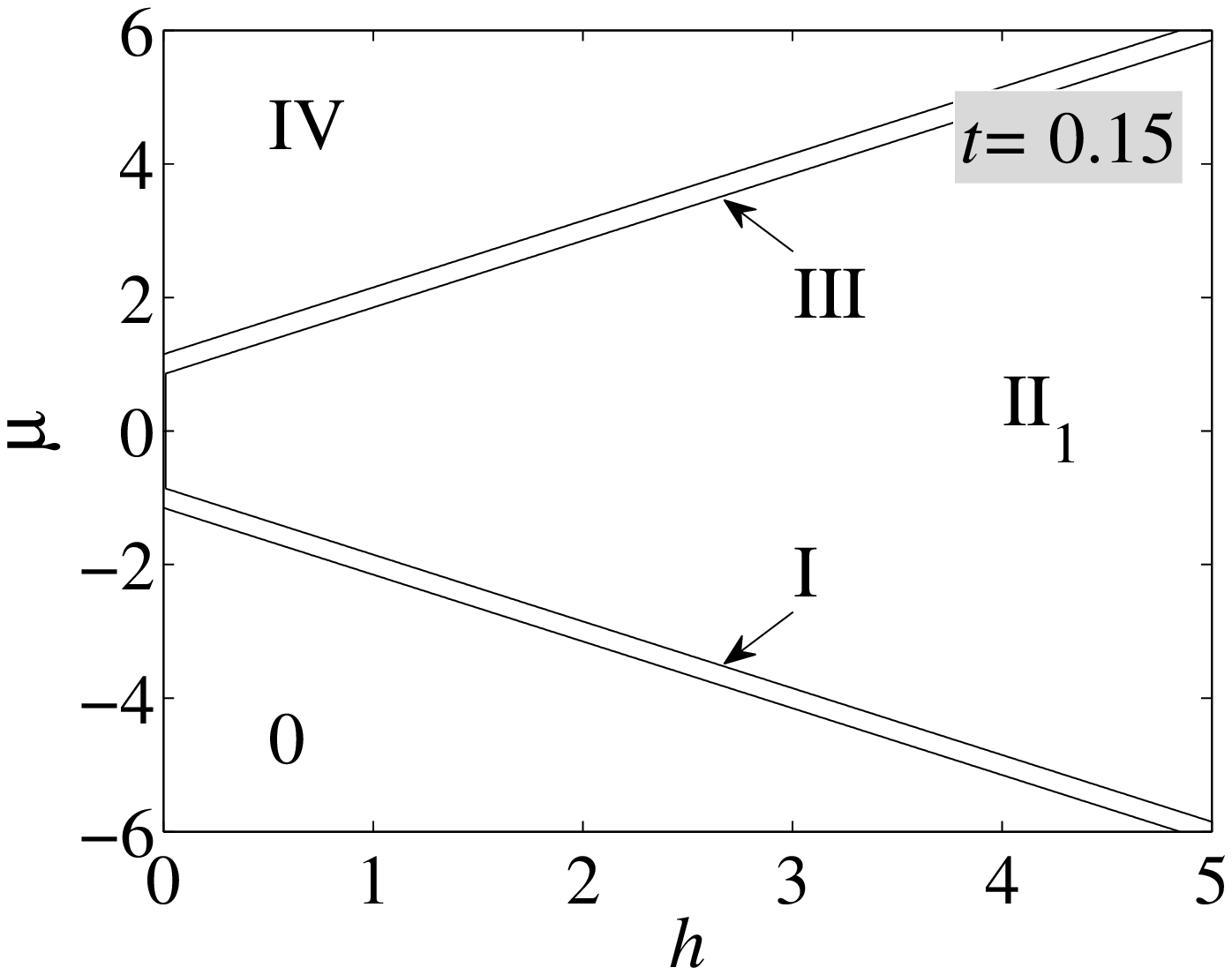}
\includegraphics[width=0.4\textwidth,trim=0 0 1.3cm 0.5cm, clip]{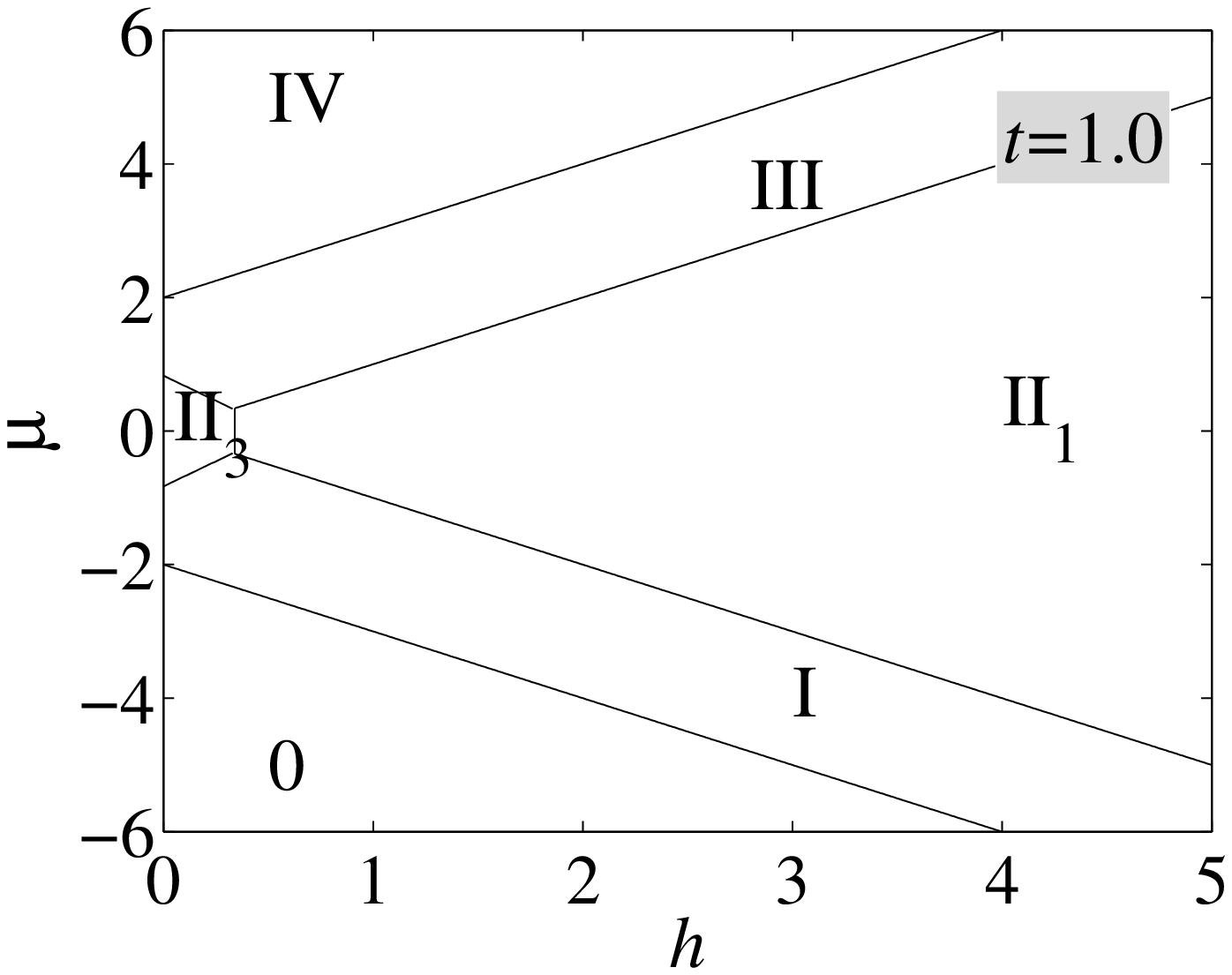}
\includegraphics[width=0.4\textwidth,trim=0 0 1.3cm 0.5cm, clip]{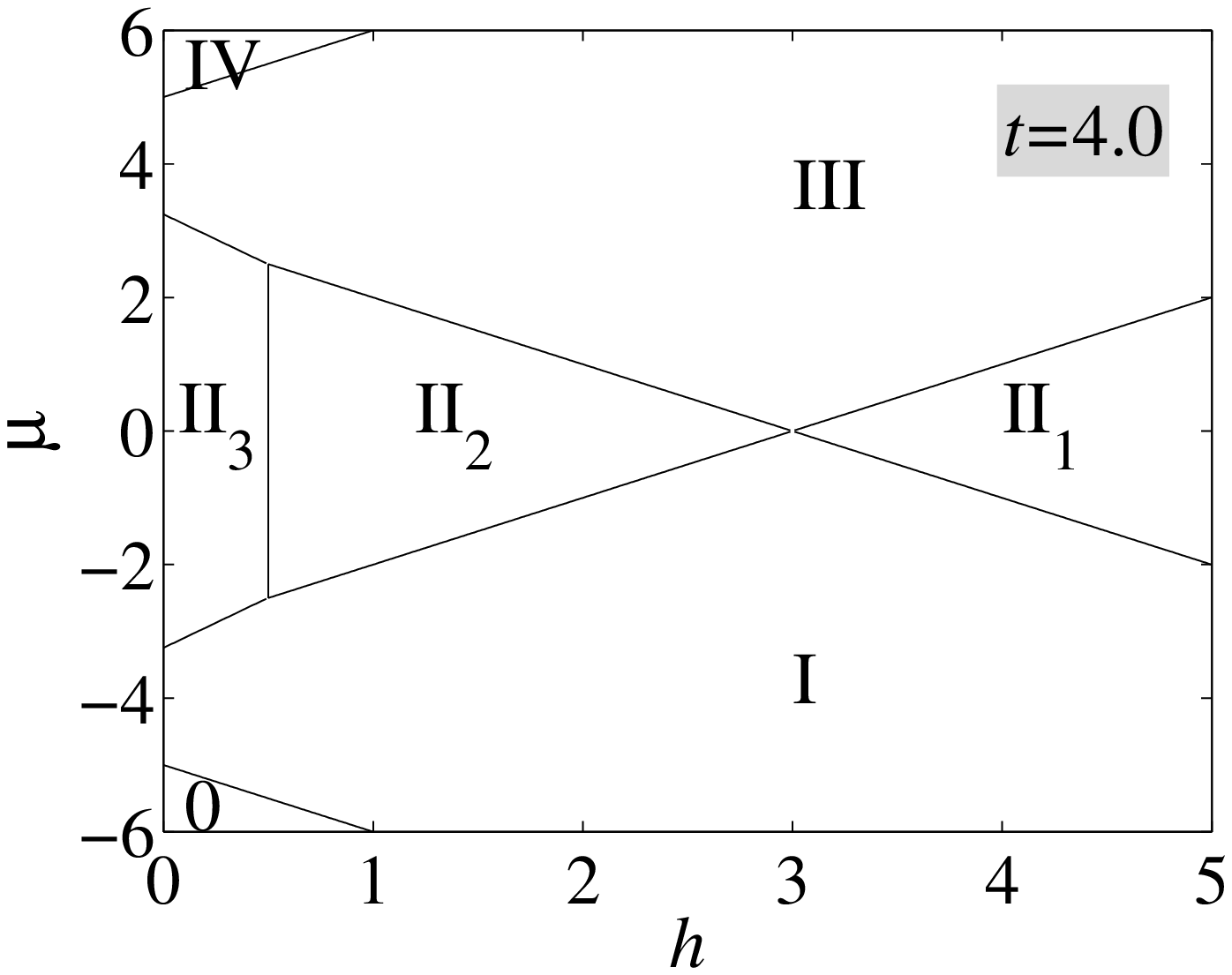}
\caption{\small Ground-state phase diagrams in the $\mu$-$h$ plane for $J=1$, $q=4$ and three selected values of $t=0.15$, $1$, and $4$.}
\label{fig2}
\end{center}
\end{figure}

After excluding the two trivial phases of the bond subsystem with $\rho=0$ (zero electron occupancy) and $\rho=4$ (full-filling), we identify three types of ground-state phase diagrams  with typical examples presented in Fig.~\ref{fig2}.  
The first type, which is represented in Fig.~\ref{fig2} for $t=0.15$, occurs whenever the hopping term is below the critical value $t_c({\rm II}_1{-\rm II}_3)=\sqrt{h(1+1/q)[2J+h(1+1/q)]}$. Under this condition, the fully polarized (F) spin-electron state is present. Surprising existence of the F state within the half-filled case, for which the quantum antiferromagnetic (AF) state at $h=0$ with a perfect N\'eel order of the nodal spins was previously detected~\cite{Cenci}, can be explained as follows: rather weak correlations between the mobile electrons induced by the hopping term $t$ become insignificant in comparison with the exchange interaction $J$ between the spin and electron subsystems. Hence, it follows that an arbitrary magnetic field $h\neq0$ aligns all the spins into field direction. The minimal effect of the hopping term is likewise reflected in stabilizing the phase II$_1$ with two mobile electrons per bond, which is dominant in the phase diagram, contrary to the phases I and III with odd number of the mobile electrons per bond existing in narrow regions only. 
 
\begin{figure}[!]
\begin{center}
\includegraphics[width=0.45\textwidth,trim=0 0 1.3cm 0.5cm, clip]{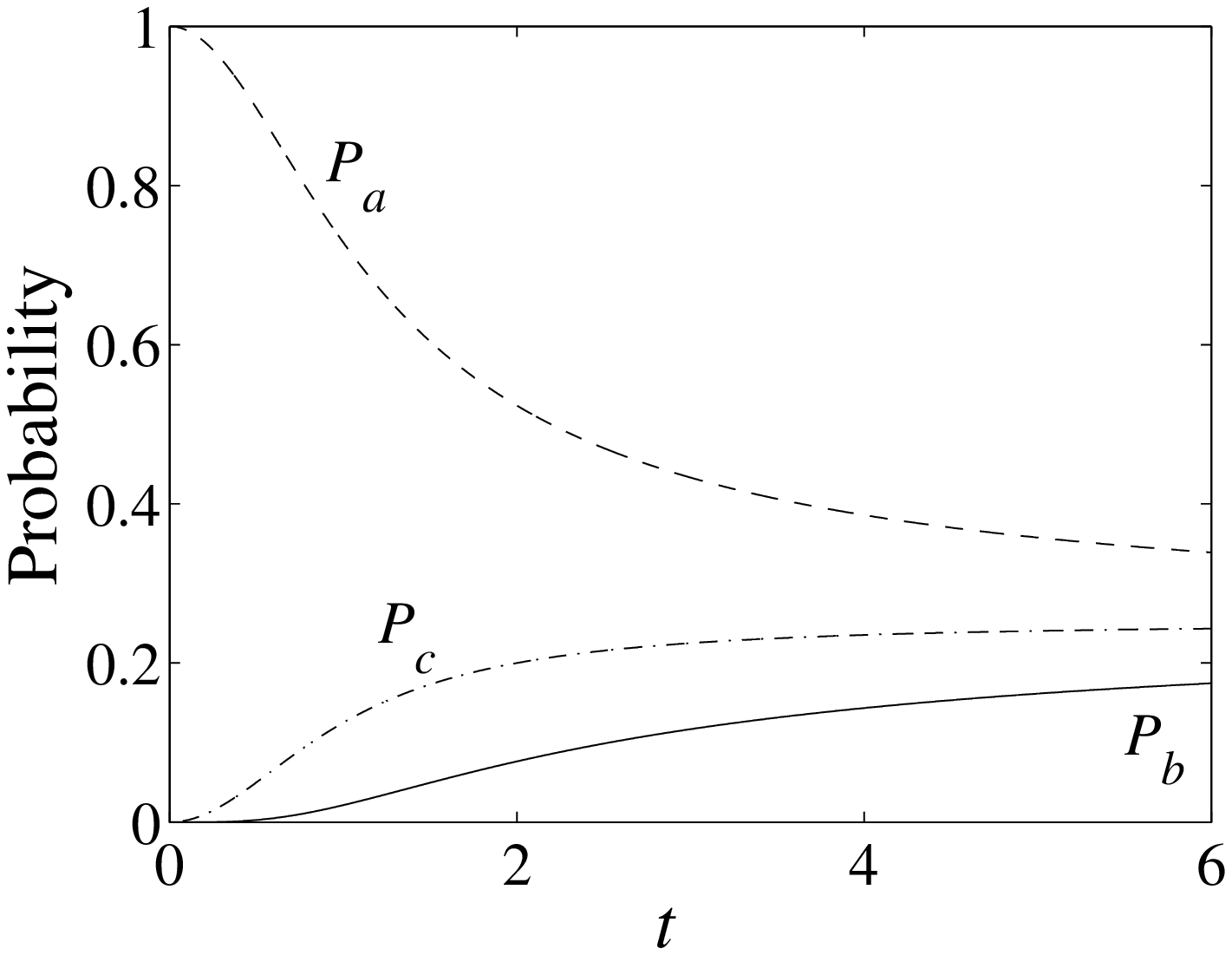}\\\hspace*{0.5cm}
{\includegraphics[width=0.4\textwidth,trim=0 0.0cm 0.cm 0.cm, clip]{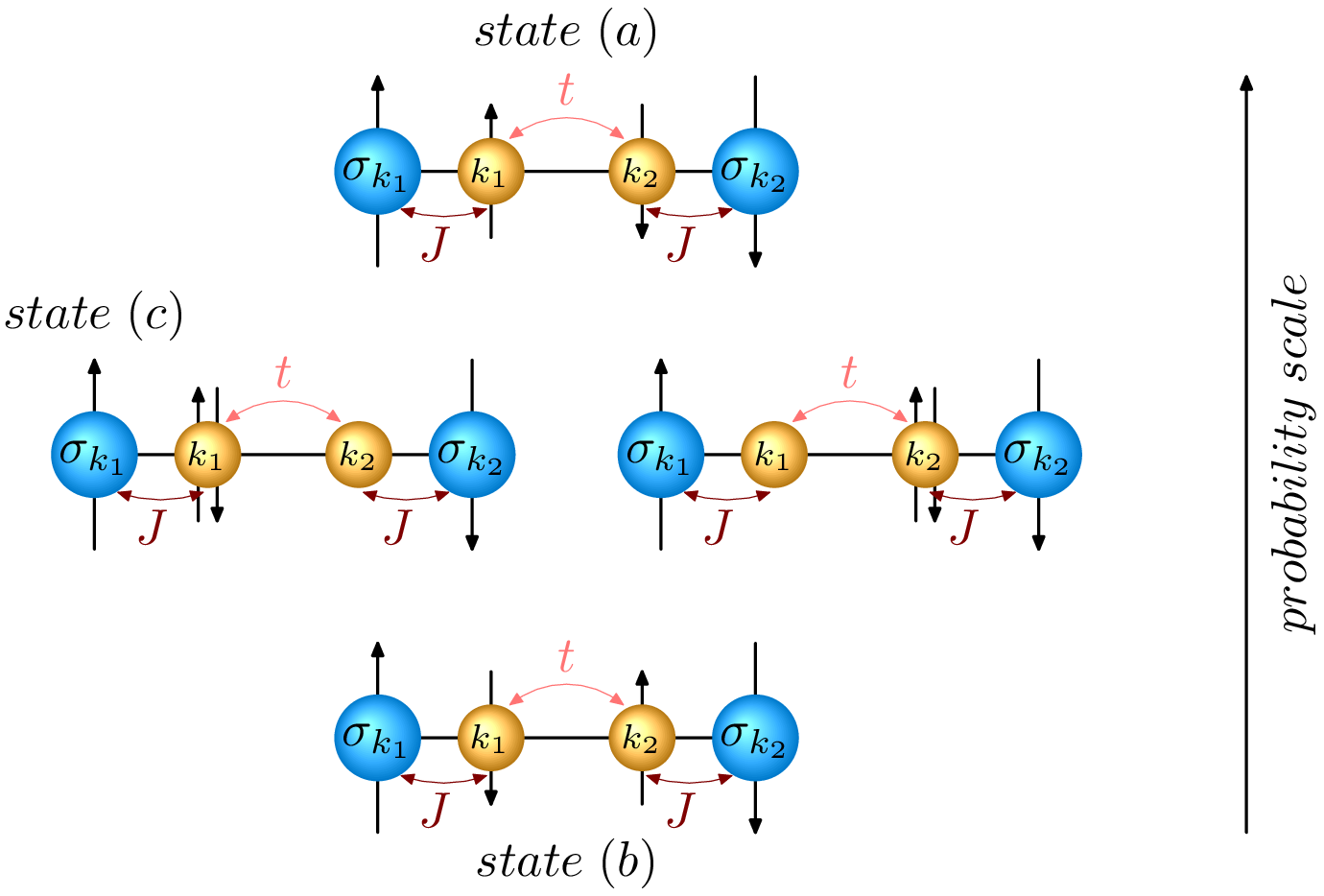}}
\caption{\small The occurrence probabilities of microstates within the ground state II$_3$ (see Tab.~\ref{tab1}), where $P_a$ determines the probability of the microstate $|\!\uparrow,\downarrow\rangle_k$ with the corresponding probability amplitude $a=\frac{J+\sqrt{J^2+t^2}}{2\sqrt{J^2+t^2}}$, $P_b$ stands for the probability of the microstate $|\!\downarrow,\uparrow\rangle_k$ with the corresponding probability amplitude $b=\frac{-(\sqrt{J^2+t^2}-J)}{2\sqrt{J^2+t^2}}$ and $P_c$ denotes the probability of the microstates $|\!\uparrow\downarrow,0\rangle_k$ and $|0,\uparrow\downarrow\rangle_k$ with the corresponding probability amplitude $c=\frac{t}{2\sqrt{J^2+t^2}}$. In numerical calculations the parameter $J$ has been set to unity.}
\label{fig3}
\end{center}
\end{figure}

The competition  among the hopping term $t$, the exchange coupling $J$, and the magnetic field $h$ becomes more intricate above the critical value of the hopping term $t>t_c({\rm II}_1{-\rm II}_3)$. Then, the quantum AF state II$_3$ (with the N\'eel spin order), which is observable for $\rho=2$ and $h=0$, persists whenever the magnetic field  is smaller than $q(\sqrt{J^2+t^2}-J)/(1+q)$. At this value, the effect of the hopping term $t$ is completely suppressed by the magnetic field, and the system undergoes a discontinuous phase transition to the F state II$_1$ (see the case $t=1$ in Fig.~\ref{fig2}). It is evident from Tab.{\ref{tab1}} that the occurrence probabilities of microstates emerging within the phase II$_3$ strongly depend on the parameters $t$ and $J$, but they do not depend on the magnetic field $h$. In addition, Fig.~\ref{fig3} demonstrates that the most probable spin orientation of the mobile electrons follows the spin orientation of the localized Ising spins, i.e. the occurrence probability of the microstate $|\!\!\uparrow,\downarrow\rangle$ is always the highest within the phase II$_3$. 
Another interesting observation is that stronger correlations between the mobile electrons stabilize the phases I and III with odd number of the mobile electrons. To conclude, the second type of the ground-state phase diagram includes six different ground-state phases and can be found for moderate values of the hopping term $t_c({\rm II}_1{-\rm II}_3)<t<t_c({\rm II}_2{-\rm II}_3)=q[J^2-(h/q)^2]/2h$. 

Last but not least, the competition between the model parameters may generate an extra phase II$_2$ at the half-filled band case whenever the hopping term exceeds the critical value $t_c({\rm II}_2{-\rm II}_3)$. The third type of the phase diagram thus totally involves seven ground states, as represented by the special case $t=4$ in Fig.~\ref{fig2}. The novel phase occurs in between the quantum AF phase II$_3$ and the classical F phase II$_1$. It can be regarded as an intermediate phase with the mixed character of F-AF order. Namely, the external magnetic field primarily forces the localized Ising spins to align into a magnetic field, but the electronic subsystem still displays a quantum AF order owing to a strong electron correlation mediated by the hopping term $t$. The phase boundaries among all the three phases II$_1$, II$_2$, and II$_3$ with two mobile electrons per bond are given by the following conditions:
\begin{align}
\mbox{II$_1$-II$_2$}&:\;h=t-J\label{eq22a},\\
\mbox{II$_1$-II$_3$}&:\;h=q(\sqrt{J^2+t^2}-J)/(1+q) \label{eq22b},\\
\mbox{II$_2$-II$_3$}&:\;h=q(\sqrt{J^2+t^2}-t). 
\label{eq22c}
\end{align}
Moreover, it can be  observed from Fig.~\ref{fig2} that  the stability regions of the phases I and III with odd number of mobile electrons per bond are in the third type of the ground-state phase diagram repeatedly wider in comparison with two aforementioned cases, which leads to the conclusion that the quantum-mechanical hopping energetically favors the configurations with odd number of mobile electrons per bond. For the sake of completeness, analytical expressions for the other ground-state phase boundaries associated with discontinuous (first-order) phase transitions are derived by comparing energies from Tab.~\ref{tab1} resulting in
\begin{align}
\mbox{0-I}&:\;\mu=-J-h-t, \nonumber\\
\mbox{I-II$_1$}&:\;\mu=-J-h+t, \nonumber\\
\mbox{I-II$_2$}&:\;\mu=J+h-t, \nonumber\\
\label{eq23}
\mbox{I-II$_3$}&:\;\mu=J+h+2h/q+t-2\sqrt{J^2+t^2}, \\
\mbox{III-II$_1$}&:\;\mu=J+h-t, \nonumber\\
\mbox{III-II$_2$}&:\;\mu=-J-h+t, \nonumber\\
\mbox{III-II$_3$}&:\;\mu=-J-h-2h/q-t+2\sqrt{J^2+t^2}, \nonumber\\
\mbox{III-IV}\,\,&:\; \mu=J+h+t.\nonumber
\end{align}
\begin{figure}[b!]
\begin{center}
{\includegraphics[width=0.35\textwidth,trim=0 0 1.3cm 0.0cm, clip]{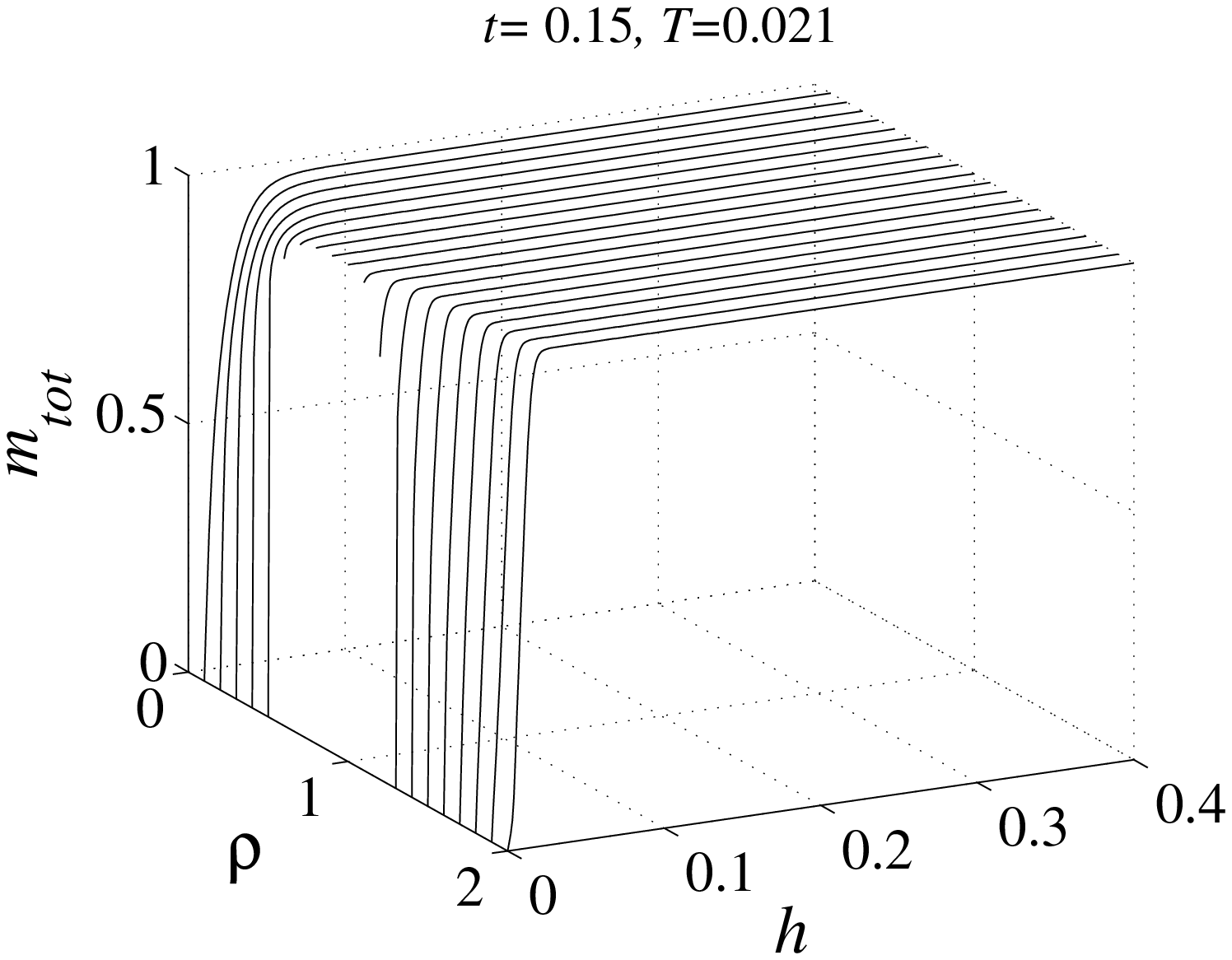}}
{\includegraphics[width=0.35\textwidth,trim=0 0 1.3cm 0.0cm, clip]{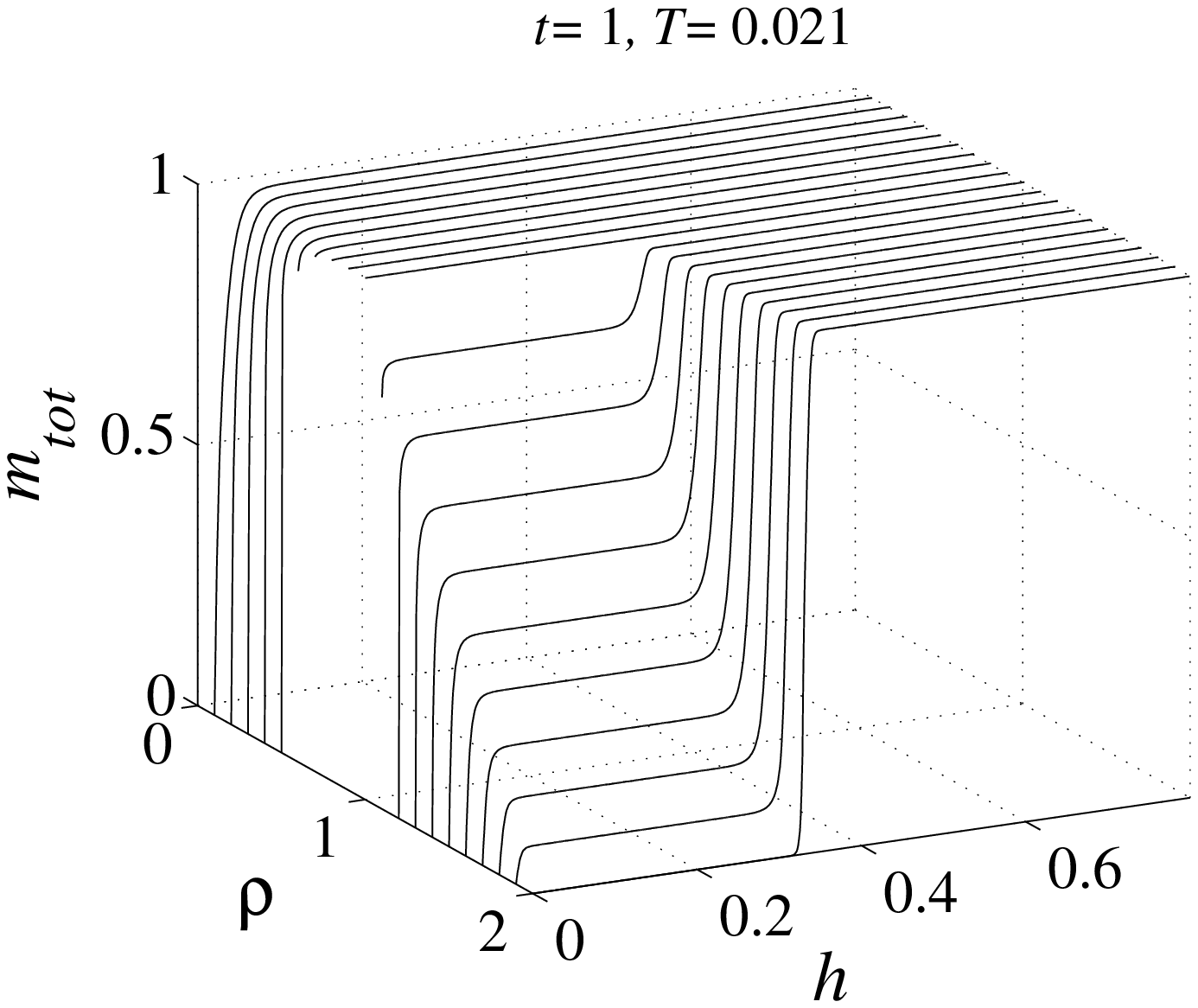}}
{\includegraphics[width=0.35\textwidth,trim=0 0 1.3cm 0.0cm, clip]{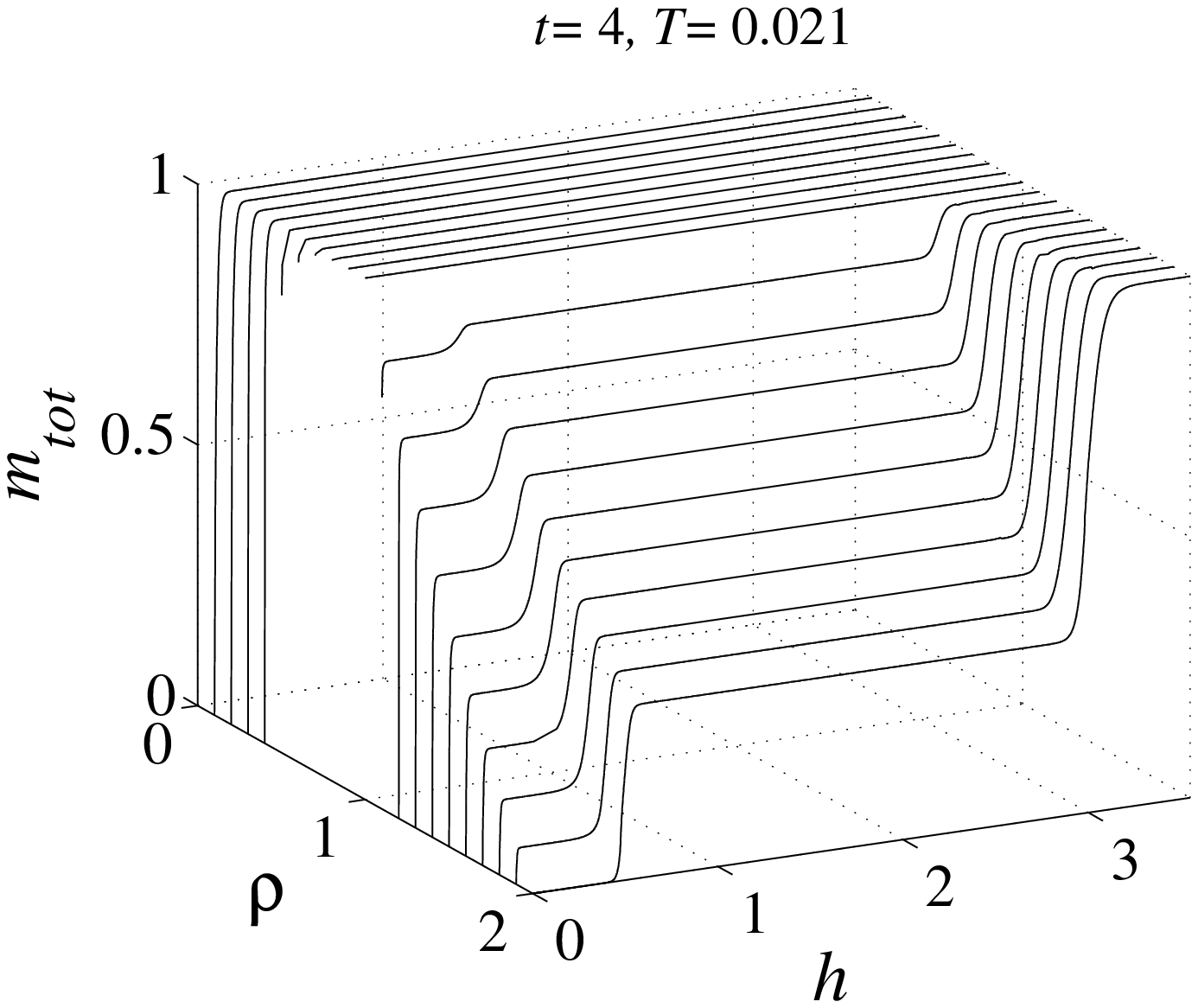}}
\caption{\small 3D plot of the total magnetization normalized with respect to its saturation value as a function of the electron density $\rho$ and the magnetic field $h$ for $J=1$, $q=4$ and three selected values of the hopping term $t$ at low temperature $T=0.021$.}
\label{fig4}
\end{center}
\end{figure}

To analyze possible metamagnetic transitions caused by the external magnetic field $h$ and the electron density $\rho$, the total uniform magnetization $m_{tot}$ of the spin-electron model on the doubly decorated square lattice is plotted in Fig.~\ref{fig4} at low temperature $T=0.021$. In general, one detects a close coincidence between the low-temperature magnetization curves shown in Fig.~\ref{fig4} and the ground-state phase diagrams depicted in Fig.~\ref{fig2}. As a matter of fact, the total magnetization $m_{tot}$ normalized with respect to its saturation value always reaches unity (excluding the zero-field case $h=0$) for $t=0.15$, which corroborates F order within both subsystems for all $0<\rho\leq 2$. The total magnetization saturates upon strengthening the magnetic field at low electron densities $0\leq \rho\leq 1$ assuming $t=1$. At the same time, it exhibits an intermediate plateau before being saturated at the higher electron densities $1<\rho\leq 2$. The stability of this intermediate plateau (in the whole range of the electron concentration $1< \rho\leq 2$) coincides with the phase II$_3$, for which the critical field $h_c$ is given by Eq.~(\ref{eq22b}). The observed magnetization plateau indeed turns into a zero magnetization plateau in the half-filling case $\rho=2$. It agrees with the AF nature of the phase II$_3$, whereas the critical field $h_c$ shows only a small shift towards lower magnetic fields upon decreasing of the electron concentration $\rho$. It is noteworthy that the height of the intermediate magnetization plateau can be continuously tuned according to the formula $m_{tot}=2(2-\rho)/(1+\rho)$ upon varying of the electron density within the range $1< \rho\leq 2$. The striking dependence of the height of intermediate plateau on the electron density can be attributed to a competition between the local F order supported by a single hopping electron per bond and the local AF order supported by a hopping of two mobile electrons per bond. Owing to this fact, the height of intermediate magnetization plateau interpolates between  $m_{tot}=1$ and $m_{tot}=0$ when the electron density changes from a quarter filling to a half filling.

The ground-state phase diagram shown in Fig. \ref{fig2} suggests that the magnetization curves at $t=4$ should include two intermediate magnetization plateaus for the electron densities $\rho>1$ in concordance with the 3D magnetization plot displayed in Fig.~\ref{fig4}. The first intermediate plateau emergent at lower magnetic fields at $m_{tot}=2(2-\rho)/(1+\rho)$ has the same origin, as described above for the case with a moderately strong hopping term,  while the second intermediate plateau originates from the mutual competition between the magnetic field and the hopping term. In the latter case, the external magnetic field is strong enough to polarize the localized Ising spins, although it does not suffice to break the AF correlation of two mobile electrons supported by the quantum-mechanical hopping process. In fact, the sublattice magnetization of the spin subsystem $m_i=1$ is saturated within this intermediate magnetization plateau unlike the sublattice magnetization of the electron subsystem depending on the electron concentration according to the formula $m_e=(2-\rho)$. Owing to this fact, the other magnetization plateau appears at the following value of the total magnetization $m_{tot}=(3-\rho)/(1+\rho)$. The critical magnetic fields, at which the investigated spin-electron system undergoes steep changes of the magnetization connected to the appearance of the phase II$_2$, coincide with the critical fields given by Eqs.~(\ref{eq22a}) and (\ref{eq22c}).

For completeness, we also investigate the staggered magnetization as an order parameter for the AF type of ordering. It turns out that the staggered magnetization is non-zero for $t>t_c({\rm II}_1{-\rm II}_3)$ and $\rho\to 2$ in accordance with the stability region of the phase II$_3$. It was found, that the non-zero value of staggered magnetization $m^s_{tot}$ is accompanied within  the phase II$_3$ by the non-zero value of the uniform magnetization $m_{tot}$ in response to a presence of magnetic field, and thus a new type of the AF ordering with F features (AF$^*$) is formed. In contrast to the uniform magnetization $m_{tot}$, the staggered magnetization $m^s_{tot}$ does not exhibit any stepwise dependence on magnetic field. Instead, it exhibits a plateau, whose height strongly depends on the electron concentration $\rho$. Here we notice that the non-zero staggered magnetization $m^s_{tot}\neq 0$ is observed at lower electron fillings in comparison with the zero-field limit $h=0$. This is an indication for existence of reentrant phase transitions. In addition, the external magnetic field can also cause the reentrant phase transitions while fixing the electron concentration $\rho$. Hence, one of the following two sequences of the reentrant transitions is generated: F$_{m_{tot}=a}$--AF$^*$--F$_{m_{tot}=1}$ or F$_{m_{tot}=a}$--AF$^*$--F$_{m_{tot}=b}$--F$_{m_{tot}=1}$ for $a<b<1$ depending on the strength of the hopping term $t$. Except the aforementioned reentrances, the investigated spin-electron model also exhibits additional field-induced phase transitions for the electron concentration close to a quarter and half filling:
\begin{align}
\begin{array}{rll}
\mbox{F}_{m_{tot}\neq 1}\mbox{--F}_{m_{tot}=1} & \mbox{if}& t_c({\rm II}_1{-\rm II}_3) < t < t_c({\rm II}_2{-\rm II}_3), \\& &\rho\to 1,\\
\mbox{F}_{m_{tot}=a}\mbox{--F}_{m_{tot}=b}\mbox{--F}_{m_{tot}=1} & \mbox{if}& t>t_c({\rm II}_2{-\rm II}_3),\\ && \rho\to 1, \\
\mbox{AF$^*$--F}_{m_{tot}=1}& \mbox{if}& t_c({\rm II}_1{-\rm II}_3) < t < t_c({\rm II}_2{-\rm II}_3),\\ && \rho\to2, \\
\mbox{AF$^*$--F}_{m_{tot}\neq 1}\mbox{--F}_{m_{tot}=1} & \mbox{if}&t>t_c({\rm II}_2{-\rm II}_3),\\ && \rho\to2,\\ 
\mbox{AF--F}_{m_{tot}=1}& \mbox{if}& t_c({\rm II}_1{-\rm II}_3) < t < t_c({\rm II}_2{-\rm II}_3),\\ && \rho=2, \\
\mbox{AF--F}_{m_{tot}\neq 1}\mbox{--F}_{m_{tot}=1} & \mbox{if}&t>t_c({\rm II}_2{-\rm II}_3),\\ && \rho=2.
\end{array}
\label{eq23a}
\end{align}

\subsection{Thermodynamics}

Let us analyze magnetic behavior of the investigated system at finite temperature using the CTMRG method~\cite{Nashino}, which is designed to compute the canonical partition function $Z_{IM}=\sum\exp[-\beta{\cal H}_{IM}(R,h_{ef})]$, cf. Eq.~(\ref{eq19}), within sufficiently high numerical accuracy. We primarily focus on the cooperative phenomena originating from the competition between the kinetic term, exchange coupling, magnetic field and temperature. Provided that the hopping term $t<t_c({\rm II}_1{-\rm II}_3)$, only the F ground states are present in the finite-temperature phase diagram  regardless of the electron filling. For $t>t_c({\rm II}_1{-\rm II}_3)$, the phase diagram in the $T$-$\rho$ plane at zero magnetic field involves paramagnetic (P), AF, and F phases, as comprehensively studied in our previous work \cite{Cenci}. 
\begin{figure}[b!]
\begin{center}
{\includegraphics[width=0.45\textwidth,trim=0 0 1.3cm 0.5cm, clip]{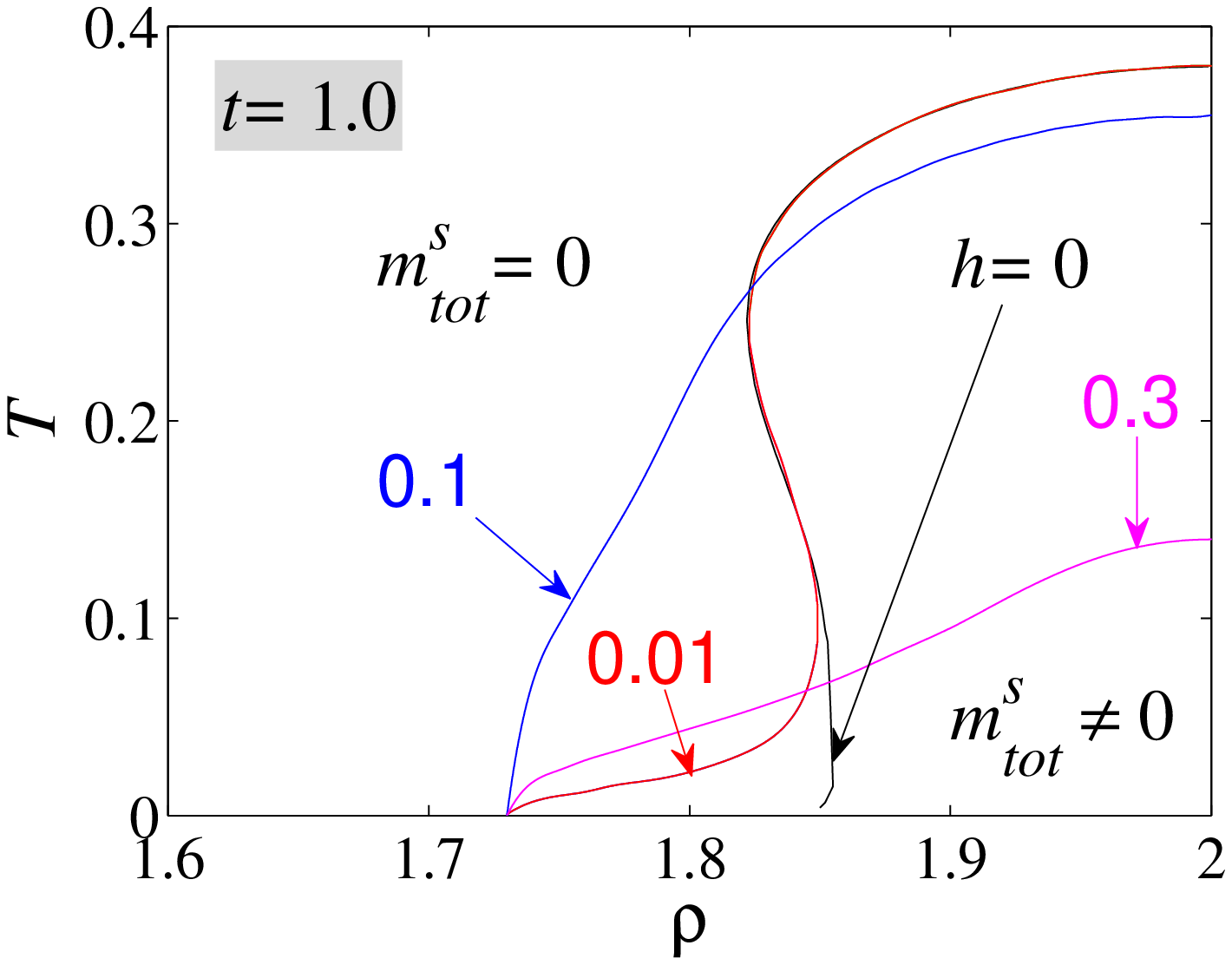}}
{\includegraphics[width=0.45\textwidth,trim=0 0 1.3cm 0.5cm, clip]{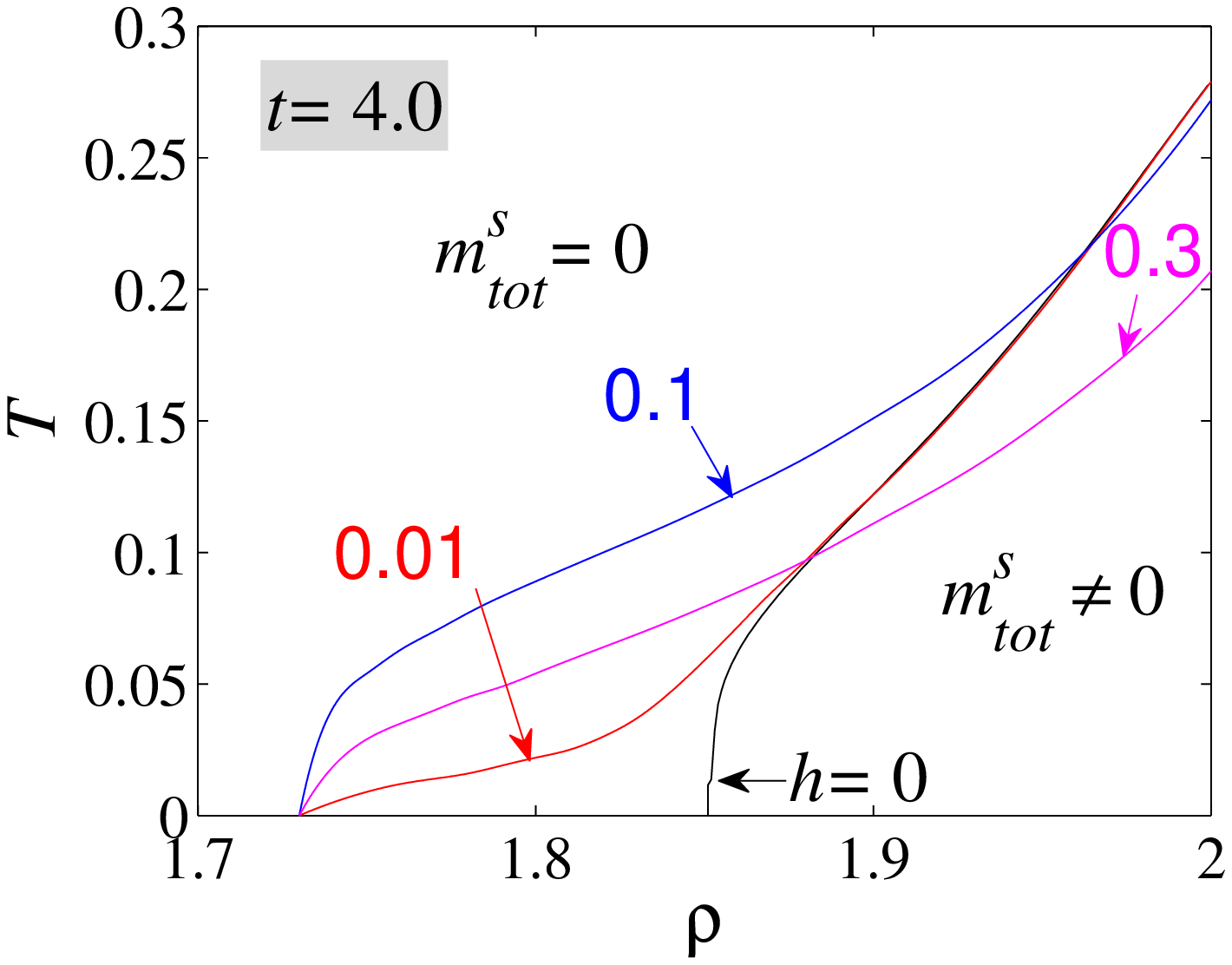}}
\caption{\small The finite-temperature phase diagrams for two representative hopping terms $t=1$ (upper panel) and $t=4$ (lower panel) for the $J=1$, $q=4$ and selected magnetic fields $h=0$, $0.01$, $0.1$, and $0.3$.}
\label{fig5}
\end{center}
\end{figure}
The effect of the external magnetic field is expected to be most pronounced within the P phase, where randomly oriented spins are forced to align in the magnetic-field direction.
Moreover, it is reasonable to assume that the AF phase shrinks within the finite-temperature phase diagram in response to strengthening of the external magnetic field. In accordance with our expectations, the critical temperature of the AF phase with $m^s_{tot}\neq 0$ reduces upon the strengthening of the magnetic field $h$ for most of the  electron concentrations. As already mentioned above, the non-zero magnetic fields generate the AF spin arrangement at slightly lower electron concentrations with respect to the zero-field case. Therefore, the phase diagram in Fig.~\ref{fig5} at $t=1$ shows interesting thermally-induced reentrant phase transitions at low magnetic fields ($h=0$ and $h=0.01$), where three consecutive phase transitions separate the sequence of the phases AF$^*$--F--AF$^*$--F for the electron concentration $\rho\approx1.84$. However, the reentrance completely vanishes at greater values of the hopping term as exemplified on the  particular case $t=4$. Besides the usual thermal reentrant phase transitions, we also plotted field-induced reentrant phase transitions at fixed non-zero temperature in Fig.~\ref{fig6}.
\begin{figure}[b!]
\begin{center}
\includegraphics[width=0.45\textwidth,trim=0 0 1.3cm 0.5cm, clip]{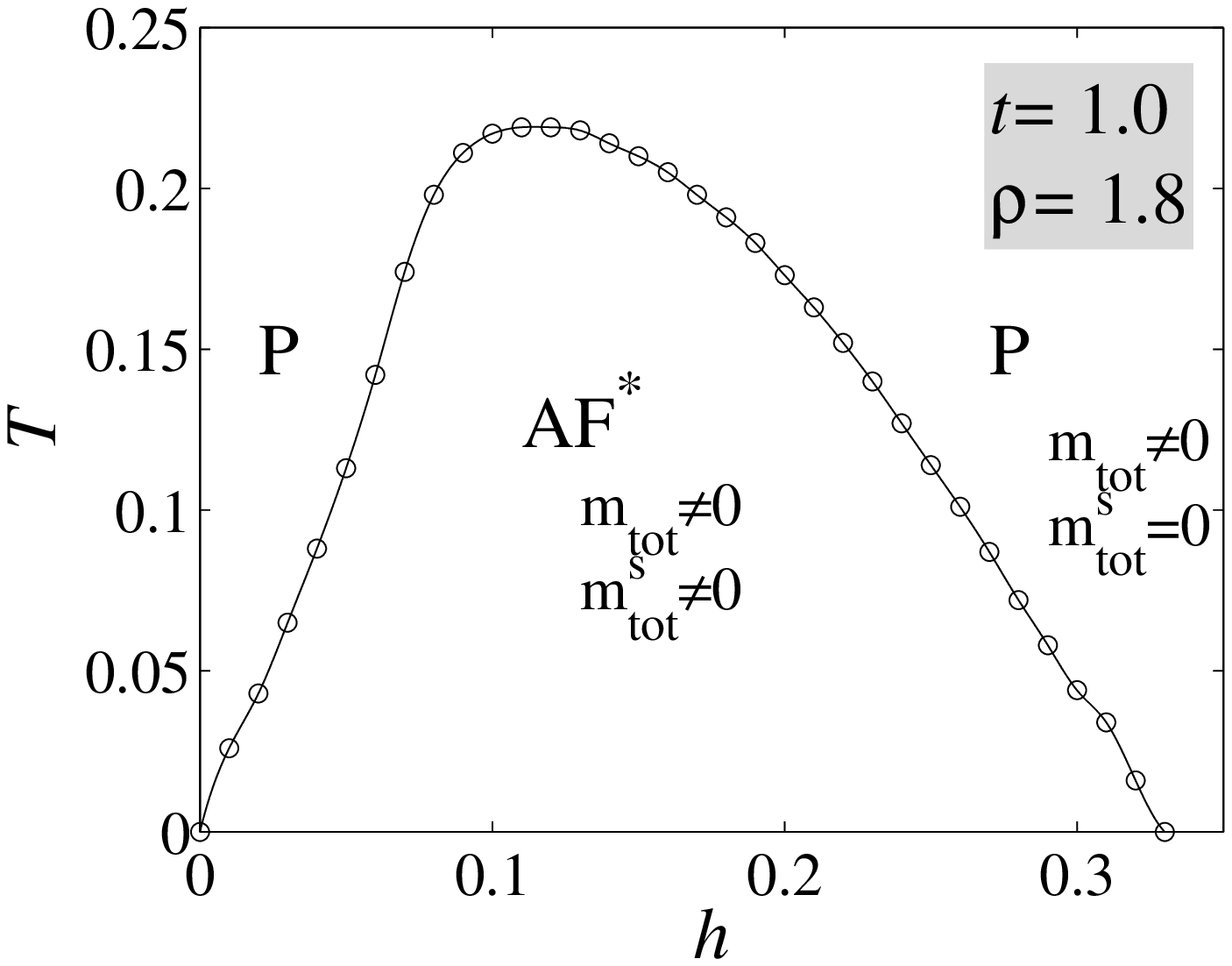}
\includegraphics[width=0.45\textwidth,trim=0 0 1.3cm 0.5cm, clip]{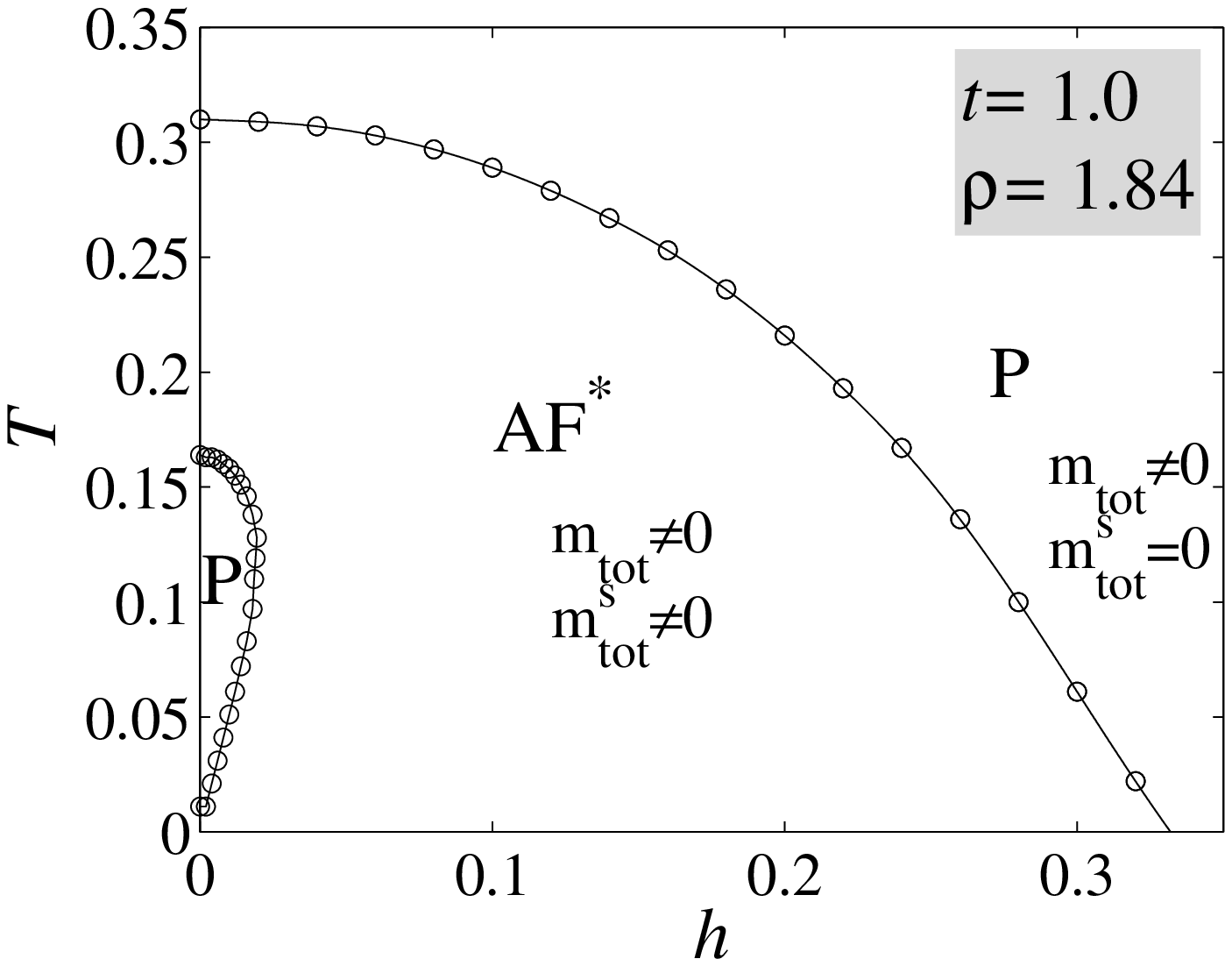}
\caption{\small The magnetic phase diagrams in the $T$-$h$ plane of the model (\ref{eq1}) for  $t=1$, $J=1$ and $q=4$. Two selected electron concentrations $\rho=1.8$ (upper panel) and $\rho=1.84$ (lower panel) illustrate the existence of field-induced reentrant phase transitions.}
\label{fig6}
\end{center}
\end{figure}
Let us recall that the AF$^*$ phase is used to denote such a parameter space, which is typical for both the non-zero  uniform $m_{tot}\neq 0$ as well as  the staggered $m_{tot}^s\neq 0$ magnetizations. The existence of the AF$^*$ phase is a consequence of the two opposite competing effects: (1) the AF order, originating from a quantum hopping process of the mobile electrons, and (2) the F order caused by the  external magnetic field. To get a deeper insight, we have analyzed thermal behavior of both sublattice magnetizations of the spin and electron subsystems along with the total magnetization. 
\begin{figure}[tb!]
{\includegraphics[width=0.45\textwidth,trim=0 0 1.1cm 0.5cm, clip]{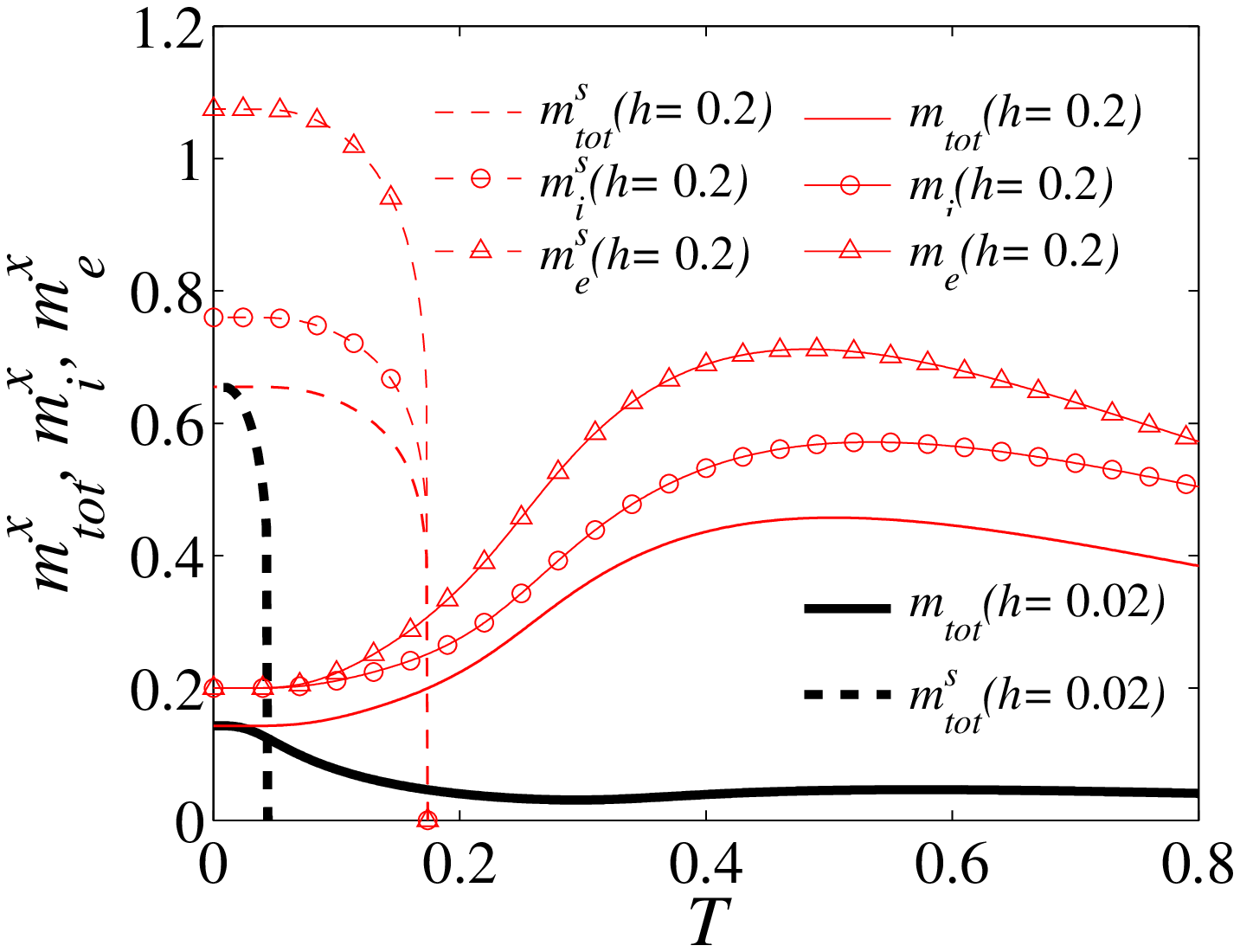}}\\
\includegraphics[width=0.45\textwidth,trim=0 0 1.1cm 0.5cm, clip]{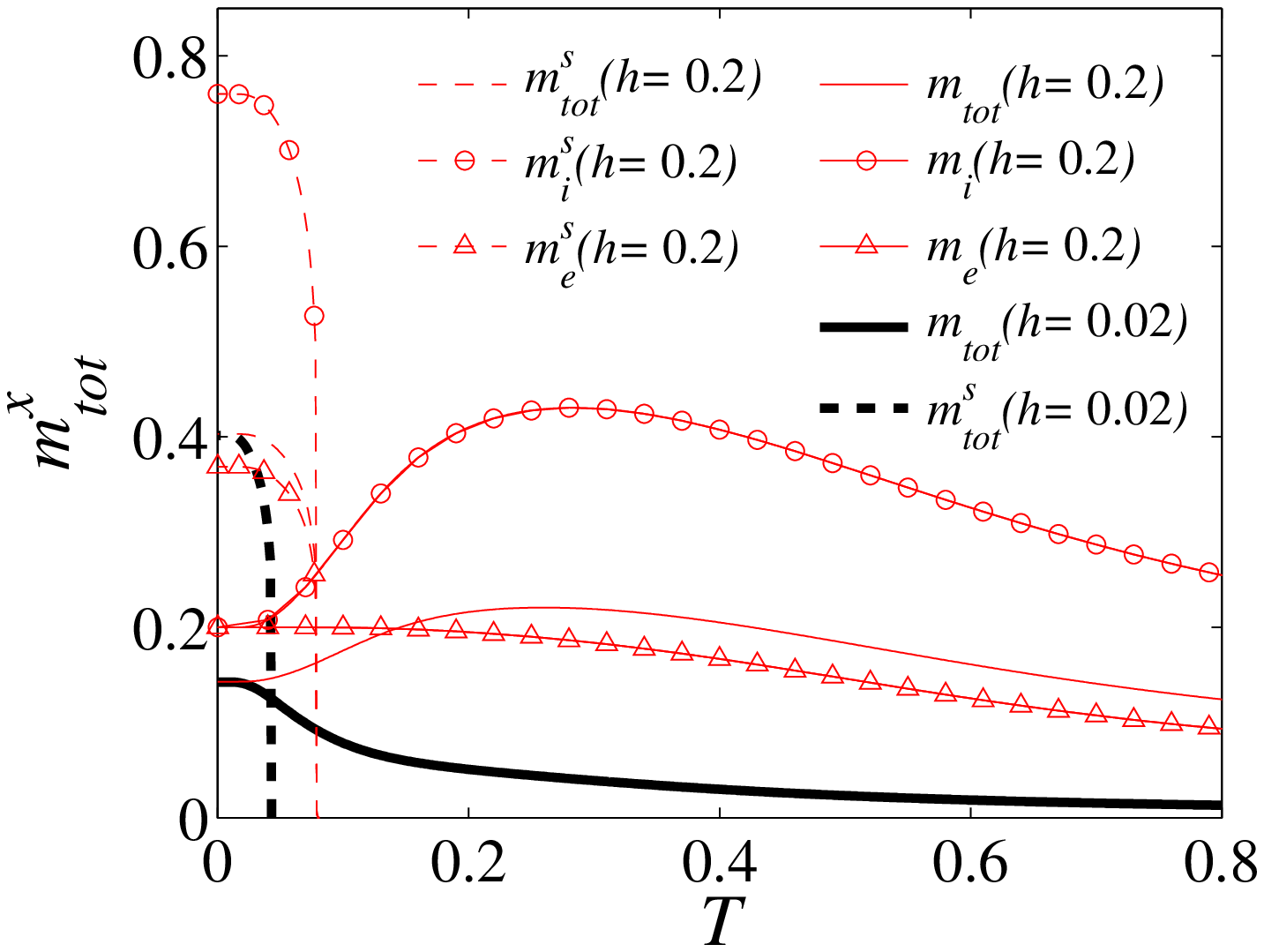}
\caption{\small The total and sublattice magnetizations as a function of temperature for the electron density $\rho=1.8$ and two selected hopping terms $t=1$ (upper panel) and $t=4$ (lower panel). Both graphs consist of data for $J=1$, $q=4$ and two distinct magnetic fields $h=0.02$ and $h=0.2$. The solid and dashed lines, respectively, illustrate the uniform $m_{tot}$, $m_i$, $m_e$ and the staggered $m_{tot}^s$, $m_i^s$, $m_e^s$ magnetizations.}
\label{fig7}
\end{figure}
The results are presented in Fig.~\ref{fig7} and Fig.~\ref{fig8}. Evidently, the existence of the AF$^*$ phase is dominantly conditioned  by the electron filling $\rho$ and, of course, by the competition of all present interactions. Both the electron filling and the value of the electron hopping $t$ strongly determine the number of AF and non-magnetic bonds in system (clearly visible from the evolution of probability in $T=0$, Fig.~\ref{fig3}), which is reflected in the value of  $m_e^s(T\to 0)$. Since the $m_i^s(T\to 0)$ is independent on $t$, the $m^s_{tot}(T\to 0)$ strongly depends on electron hopping processes. As our analyses showed, the existence of non-zero $m_{tot}$  affected by the external magnetic field $h$ is indirectly conditioned by the electron subsystem, because the lower (higher) number of AF bonds produces the smaller (stronger) damped forces 
 to reorient the  magnetic moments into the field direction. Consequently, the existence and character of the AF$^*$ phase is strongly determined by the features of electron subsystem. Another interesting observation is that the AF$^*$ phase maintains its mixed ferro-antiferromagnetic character up to higher temperatures at higher magnetic fields.
\begin{figure}[bth]
\begin{center}
{\includegraphics[width=0.45\textwidth,trim=0 0 1.1cm 0.5cm, clip]{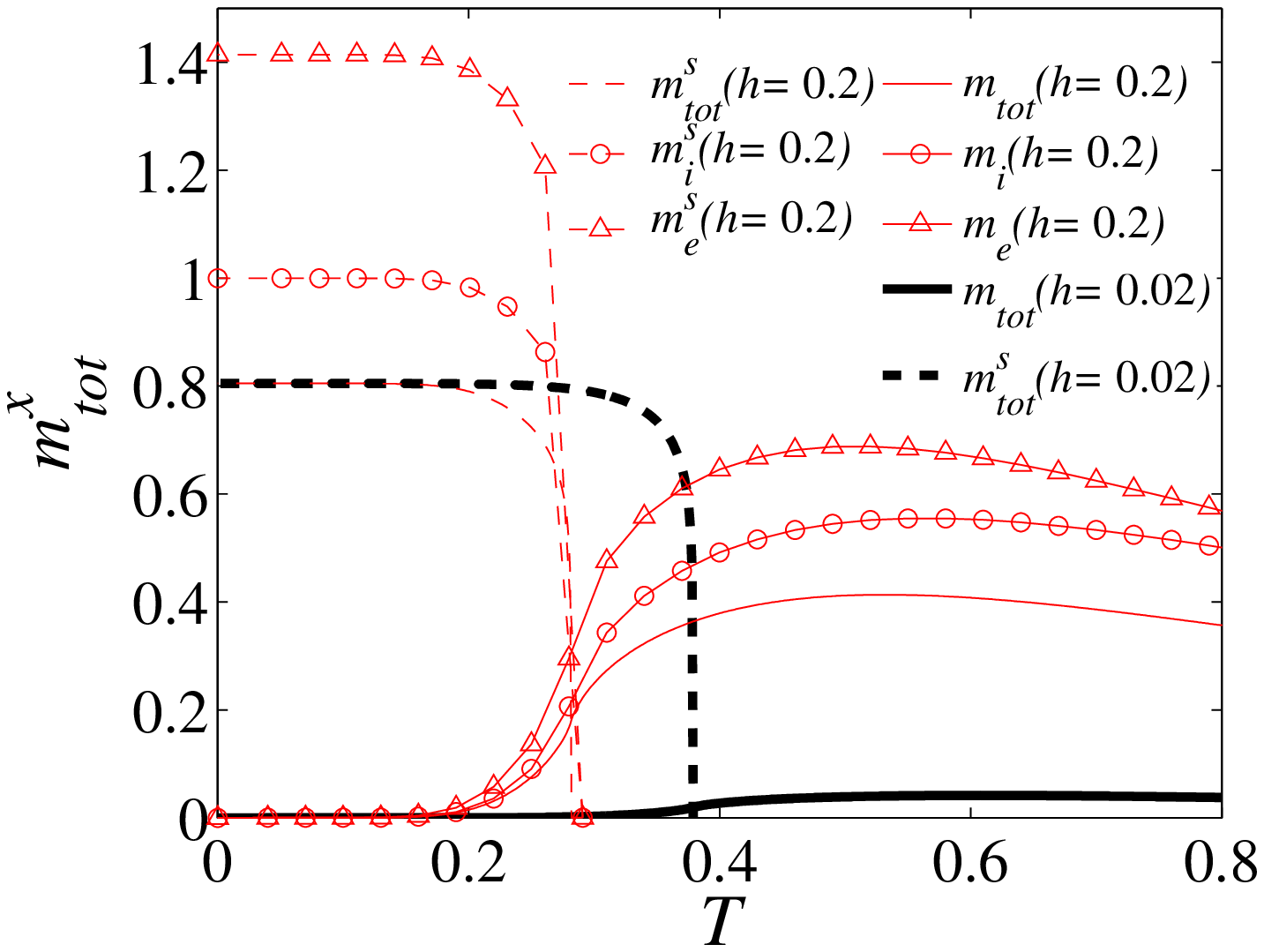}}
{\includegraphics[width=0.45\textwidth,trim=0 0 1.1cm 0.5cm, clip]{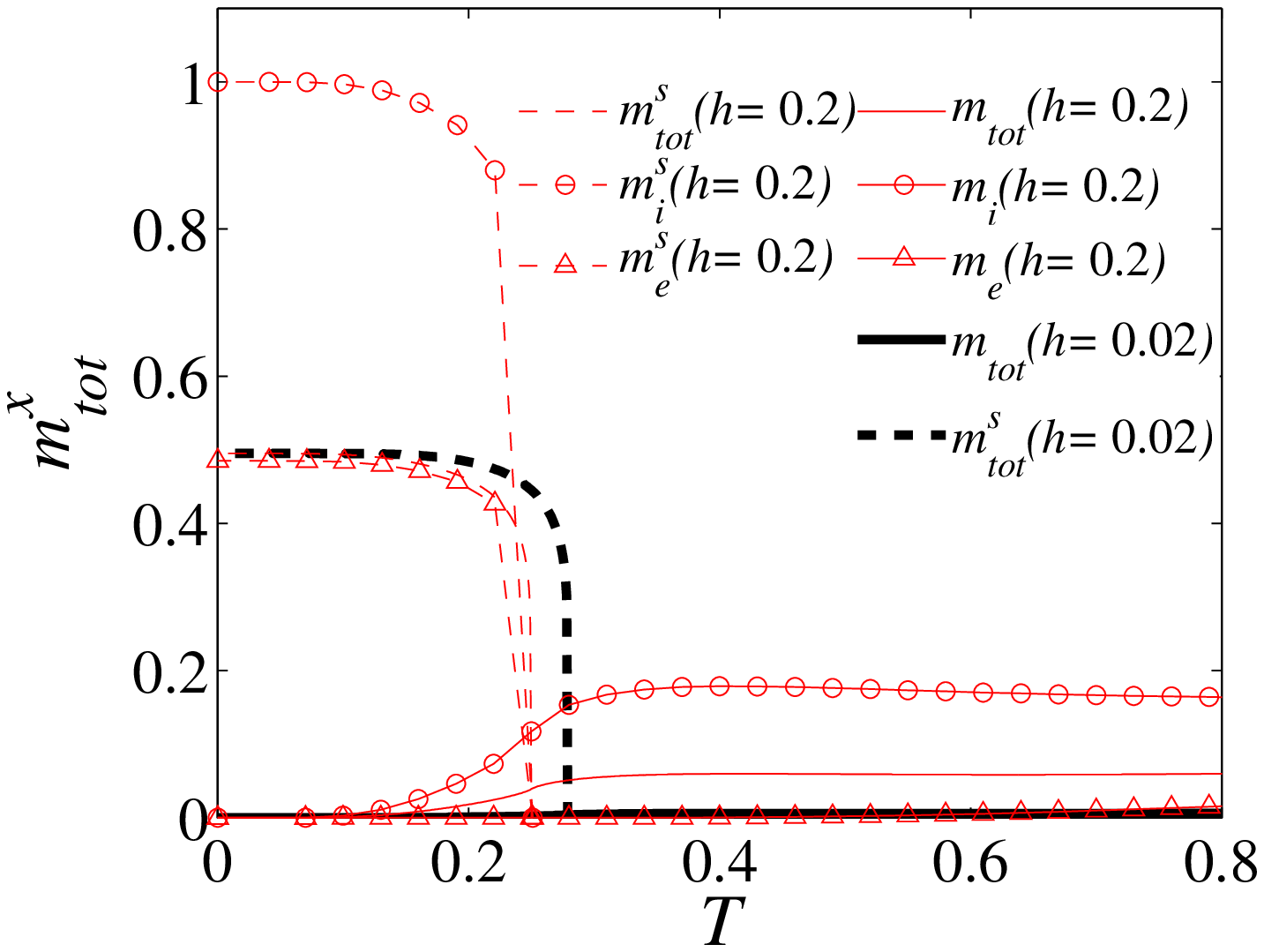}}
\caption{\small Temperature dependence of the total magnetizations $m^x_{tot}$ at $\rho=2.0$ for the two hopping terms $t=1$ (upper panel) and $t=4$ (lower panel). Both graphs consist of data for $J=1$, $q=4$ and two distinct magnetic fields $h=0.02$ and $h=0.2$. The solid and dashed lines, respectively, illustrate the uniform $m_{tot}$, $m_i$, $m_e$ and the staggered $m_{tot}^s$, $m_i^s$, $m_e^s$ magnetizations.}
\label{fig8}
\end{center}
\end{figure}

\begin{figure}[tb]
\begin{center}
{\includegraphics[width=0.45\textwidth,trim=0 0 1.1cm 0.5cm, clip]{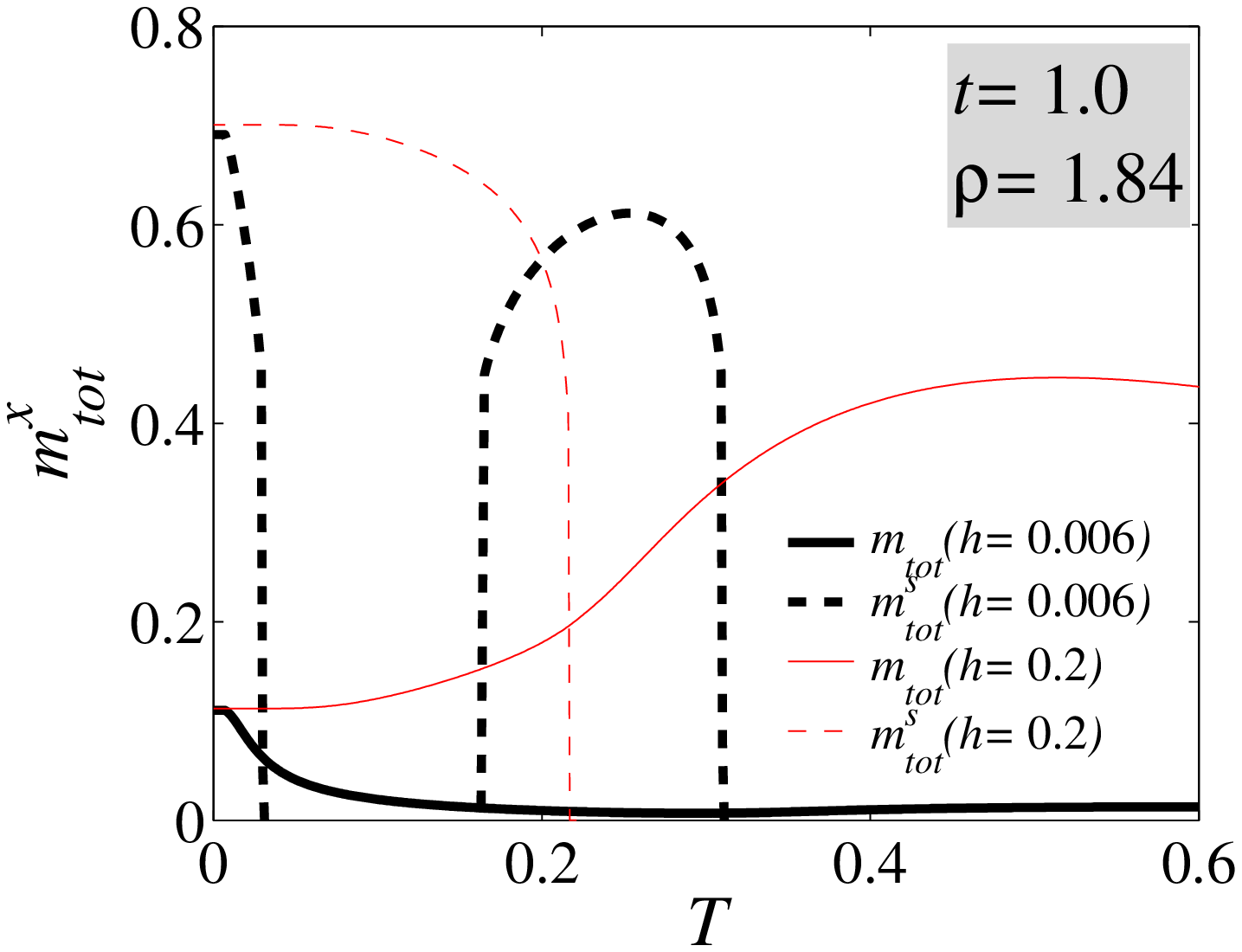}}
\caption{\small The total magnetizations against to temperature at the electron density $\rho=1.84$, for the nearest-neighbor interaction $J=1$, the coordination number $q=4$, the hopping term $t=1$ and  two distinct magnetic fields $h=0.006$ (thick black lines) and $h=0.2$ (thin red curves).}
\label{fig9}
\end{center}
\end{figure}

However, the most interesting thermal behavior of the magnetization can be detected when thermal reentrant phase transitions take place. Typical thermal variations of the magnetization with successive reentrant phase transitions are presented in Fig.~\ref{fig9} for the electron density $\rho=1.84$ and the hopping term $t=1$. 
It is obvious from Fig.~\ref{fig9} that there exist two regions with the non-zero staggered magnetization $m_{tot}^s \neq 0$ and the non-zero uniform magnetization $m_{tot} \neq 0$ for the sufficiently small magnetic field $h=0.006$. As one can see, the increasing temperature basically reduces the total uniform magnetization $m_{tot}$ as well as the total staggered magnetization $m_{tot}^s$. The latter AF order parameter $m_{tot}^s$ becomes zero at moderate temperatures, while the former F order parameter $m_{tot}$ retains non-zero albeit relatively small value due to the non-zero external magnetic field. The AF order re-appears at higher temperatures as evidenced by a sudden uprise of the total staggered magnetization $m_{tot}^s$, which finally disappears at third (highest) critical temperature. The thermal reentrance naturally vanishes at higher magnetic fields (e.g. $h=0.2$), whereas the total uniform and staggered magnetizations $m_{tot}$ and $m_{tot}^s$ then become almost thermally independent at low enough temperatures.
 
Our thermal analysis can imply a great potential of the studied spin-electron system for technological applications, because different magnetic states are controllable by various external parameters, such as temperature, magnetic field and/or electron density. To summarize, one may tune the investigated spin-electron model across several types of the magnetic phase transitions with respect to the electron concentration:
\begin{align}
\begin{array}{ll}
\mbox{AF$^*$--F} & \mbox{$\rho$ out of thermal reentrance},\\
\mbox{AF--F} &  \mbox{$\rho=2$, small $h$}, \\
\mbox{AF--AF$^*$--F} &  \mbox{$\rho=2$, large $h$},\\
\mbox{AF$^*$--F--AF$^*$--AF--F} &  \mbox{$\rho$ at the thermal reentrance}.
\end{array}
\hspace{-1cm}\label{eq23b}
\end{align}

\section{Conclusions}
\label{Conclusions}
In the present paper we have examined the coupled spin-electron model on the doubly decorated square lattice in presence of the external magnetic field by combining the analytic decoration-iteration mapping transformation with the numerical CTMRG method. Our analysis was primarily concentrated on the magnetization processes elucidating the intermediate magnetization plateaus, metamagnetic transitions, and the reentrant phase transitions. Both the ground-state and the finite-temperature phase diagrams were studied in detail with respect to the electron filling. It has been found that a spin arrangement emerges within individual ground states and strongly depends on the mutual interplay among the hopping term, the exchange interaction, the external magnetic field, and the electron concentration. The non-zero values of external magnetic field result in the richer spectrum of magnetic ground-state phase diagrams. Three types of the ground-state phase diagrams were identified, which depend on the electron hopping term. The first type of the phase diagram solely exhibits the F type of ordering in both the spin and electronic subsystems within the entire parameter space. The remaining two types of the ground-state phase diagrams contain magnetic states with the AF ordering in the both subsystems; an even more strikingly, a combined F ordering of the localized spins accompanied with the AF ordering of the mobile electrons. These novel ground states are responsible for the appearance of the intermediate plateaus in low-temperature magnetization curves including metamagnetic transitions in between them. In addition, it has been shown that the intermediate magnetization plateaus emerge above the quarter filling ($\rho>1$) only, and the height of magnetization plateaus is continuously tunable by the electron doping as evidenced by the derived exact formulae.

The most remarkable finding refers to the  AF$^*$ phase detected close to the half-filling case $\rho \to 2$, which simultaneously carries non-zero uniform and staggered total magnetizations $m_{tot} \neq 0$ and $m_{tot}^s \neq 0$, respectively. The existence of such a phase is the direct consequence of the present magnetic field because its existence has not been determined in the zero-field counterpart yet. Moreover, it turns out that the AF$^*$ phase can re-appear at higher temperatures on account of reentrant phase transitions driven either by temperature or magnetic field. The most surprising finding is that a relatively simple spin-electron model can describe the existence of phase with the F as well as AF features along with other significant magnetic phenomena of cooperative nature, which  have been experimentally observed in several real magnetic materials. In particular, doped manganites exhibit quasi-2D character~\cite{Ganguly} and the magnetic behavior basically depending on the electron doping, whereas the AF and F orders may indeed coexist together in some manganites~\cite{Moreo}. It is also generally known that the manganites also exhibit other unconventional phenomena~\cite{Salamon,Reis}, which may originate from a competition between the localized and mobile magnetic particles. In this regard, our simple model reproduces several  magnetic features such as multistep magnetization curves, metamagnetic transitions, and reentrant phase transitions, which all arise from the mutual competition of the kinetic term, the exchange coupling, the magnetic field, and the electron density. Our theoretical achievements presented in this work thus have obvious potential to contribute significantly  explaining the unconventional cooperative phenomena of the correlated spin-electron systems. 

\vspace{0.5cm}
This work was supported by the Slovak Research and Development Agency (APVV) under Grants No. APVV-0097-12, APVV-0808-12 and APVV-16-0186. The financial support provided by the VEGA under Grants No. 1/0043/16  and 2/0130/15 is also gratefully acknowledged. 

\end{document}